\documentclass[structabstract]{aa}
\usepackage{amsmath}
\usepackage{amssymb}
\usepackage{natbib}
\bibpunct{(}{)}{;}{a}{}{,} 
\usepackage{graphicx}
\usepackage{ctable}
\usepackage{subfigure}
\usepackage{txfonts}

\newcommand{\Msun}{M_{\odot}}
\newcommand{\Rsun}{R_{\odot}}

\begin{document}

\title{Bridging the gap: disk formation in the Class 0 phase with ambipolar diffusion and Ohmic dissipation}
\titlerunning{Disks in the Class 0 phase with AD and OD}

\author{{Wolf~B.~Dapp}\inst{\ref{inst1},\ref{inst2}} \and {Shantanu~Basu}\inst{\ref{inst2}} \and {Matthew~W.~Kunz} \inst{\ref{inst3},\ref{inst4}}}

\institute{J\"{u}lich Supercomputing Centre (JSC), 
					 Institute for Advanced Simulation (IAS), FZ J\"{u}lich, 52425 J\"{u}lich, Germany, 
           \email{w.dapp@fz-juelich.de} \label{inst1} \and            
					 Department of Physics \& Astronomy, The University of Western Ontario, 1151 Richmond St., London, ON, N6A 3K7, Canada, 
           \email{basu@uwo.ca}\label{inst2} \and 					            
           Rudolf Peierls Centre for Theoretical Physics, University of Oxford, 1 Keble Road, Oxford OX1 3NP, U. K. \label{inst3} \and 
					 Einstein Postdoctoral Fellow, Department of Astrophysical Sciences, Princeton University, Peyton Hall, 4 Ivy Lane, Princeton, NJ 08544, U. S. A.,
					 \email{kunz@astro.princeton.edu}\label{inst4}
					}

\date{Received 11 August 2011 / accepted 21 February 2012}

\abstract
{Ideal magnetohydrodynamical (MHD) simulations have revealed 
catastrophic magnetic braking in the protostellar phase, which prevents
the formation of a centrifugal disk around a nascent protostar.}
{We determine if non-ideal MHD, including the effects of ambipolar 
diffusion and Ohmic dissipation determined from a detailed chemical network
model, will allow for disk formation at the earliest stages of star formation.}
{We employ the axisymmetric thin-disk approximation in order to resolve a
dynamic range of 9 orders of magnitude in length and 16 orders of magnitude in 
density, while also calculating partial ionization using up to 19 species 
in a detailed chemical equilibrium model. Magnetic braking is applied to the 
rotation using a steady-state approximation, and a barotropic relation is
used to capture the thermal evolution.}
{We resolve the formation of the first and second cores, with 
expansion waves at the periphery of each, a magnetic diffusion shock, 
and prestellar infall profiles at larger radii. Power-law profiles in 
each region can be understood analytically. After the formation of 
the second core, the centrifugal support rises rapidly and a low-mass
disk of radius $\approx 10\,R_{\odot}$ 
is formed at the earliest stage of star formation,
when the second core has mass $\sim 10^{-3}\,M_{\odot}$. The mass-to-flux
ratio is $\sim 10^4$ times the critical value in the central region.}
{A small centrifugal disk can form in the earliest stage of star
formation, due to a shut-off of magnetic braking caused by magnetic field 
dissipation in the first core region. There is enough angular momentum loss to allow the 
second collapse to occur directly, and a low-mass stellar core to form with 
a surrounding disk. 
The disk mass and size will depend upon how the angular momentum transport
mechanisms within the disk can keep up with mass infall onto the disk. 
Accounting only for direct infall, we estimate that the disk will remain $\lesssim 10~\mathrm{AU}$, undetectable even by ALMA, for $\approx 4 \times 10^{4}~\mathrm{yr}$, 
representing the early Class 0 phase. 
}

\maketitle

\section{Introduction}\label{sec:Introduction}

Near the beginning of the star formation process, prestellar cores are observed to be rotating \citep[e.g.,][]{GoldsmithArquilla1985,GoodmanEtAl1993,CaselliEtAl2002}. At the end of the process, observations show nearly all Class II objects to be surrounded by rotating disks of size $\gtrsim 100~\mathrm{AU}$, likely in centrifugal balance undergoing Keplerian rotation \citep[see, e.g.,][and references therein]{AndrewsWilliams2005,WilliamsCieza2011}. Observational data concerning what happens in between those two snapshots is relatively scarce. Currently, there is no evidence for the presence of centrifugal disks larger than $\approx 50~\mathrm{AU}$ around Class 0 or Class I objects \citep[e.g.,][]{MauryEtAl2010}. However, there are outflows observed around these young objects, commonly linked to disk accretion. Disks must therefore form with a small size and grow over time. The dearth of observational data at smaller scales will be remedied by ALMA. In the meantime, the evolution of angular momentum and the formation of a centrifugal disk have to be studied theoretically and numerically. 

In order to form stars of the observed sizes and rotation rates, most of the angular momentum has to be removed from the infalling gas. This is the classical
\textit{angular momentum problem} \citep[e.g.,][]{Spitzer1978}. The required reduction of angular momentum is often credited to \textit{magnetic braking}, which acts during the contraction and collapse by linking the core with its envelope and transferring angular momentum \citep[see][and references therein]{BasuMouschovias1994}. Recent numerical simulations of protostellar collapse under the assumption of ideal magnetohydrodynamics (MHD) have suggested the converse problem, namely that cores may experience \textit{catastrophic magnetic braking}.
This is the result that an extremely pinched magnetic field 
configuration has enough strength and exerts a long enough lever arm on the 
inner regions of collapse that magnetic braking can suppress the formation 
of a centrifugal disk entirely.

Catastrophic magnetic braking was first demonstrated by 
\citet{AllenLiShu2003}, who used two-dimensional (2D) axisymmetric 
calculations. Subsequent ideal MHD simulations by \citet{MellonLi2008} 
in 2D and \citet{HennebelleFromang2008} in three dimensions (3D) showed 
that catastrophic magnetic braking occurs for initially 
aligned (magnetic field parallel to rotation axis) rotators in which the 
magnetic field is strong enough that $\mu \lesssim 10$, where $\mu$ 
is the initial core mass-to-flux ratio normalized to its critical value 
for gravitational collapse. Disk formation in ideal MHD then requires
at least $\mu \gtrsim 10$, which is significantly greater
than the typical value $\sim 2$ for dense cores determined from observations 
\citep[e.g.,][]{TrolandCrutcher2008}.
Furthermore, \citet{MellonLi2009} found that 
the inclusion of non-ideal effects of Ohmic dissipation and 
ambipolar diffusion in a 2D model was not sufficient to enable disk 
formation on scales of $\gtrsim 10$ AU from the center that were 
resolvable in their simulation. In a newer 2D model that also includes the 
Hall term of non-ideal MHD, \citet{LiKrasnopolskyShang2011} again find 
essentially the same result. All of the above simulations have an inner 
resolution limit of $\approx 10$ AU due to the use of a sink cell or an
artificially stiff equation of state at high densities. Since these simulations 
also terminated soon after the central stellar core formed, 
the main common result of these simulations is that large $\sim 100$ AU 
disks (as observed around Class II objects) cannot form during the 
early Class 0 phase. A 3D ideal MHD model by 
\citet{HennebelleCiardi2009} that starts with a non-aligned rotator, 
also finds catastrophic magnetic braking on these scales but within a 
smaller range of $\mu \lesssim 3$. The effect of numerical reconnection 
may account for some of the differences amongst these models.
All of the above works leave open the possibility that a small disk may 
actually form within the inner $\sim 10$ AU during the earliest phases of 
star formation.

An interesting question is: can a very small disk form in the innermost
regions (near the protostar) in the case of ideal MHD?
\citet{GalliEtAl2006} developed a simplified analytic model with a split-monopole
magnetic field to show that catastrophic magnetic braking extends to the 
stellar surface in the flux-frozen limit. 
\citet{KrasnopolskyKoenigl2002} and \citet{BraidingWardle2011} employed 
self-similar non-ideal MHD models in the thin-disk approximation, using 
the prescription for steady-state magnetic braking developed by \citet{BasuMouschovias1994}.
Depending on different levels of effective diffusivity, they found either
catastrophically-braked or disk-formation solutions.
However, the diffusivity was parametrized in a way that preserved self-similarity, and its 
magnitude and functional form did not make contact with microphysical models.
Consequently, a prediction for whether or not disks may actually form was not possible.

Our work fills the gap of numerically studying disk formation in the 
early Class 0 phase down to scales smaller than a stellar radius. 
In a previous paper \citep[][hereafter DB10]{DappBasu2010} we followed the 
evolution of a collapsing molecular cloud core in axisymmetric thin-disk 
geometry all the way to protostellar densities ($n>10^{20}~\mathrm{cm}^{-3}$). 
We found that Ohmic dissipation alone (in a simplified form) reduces the 
magnetic field strength sufficiently to render magnetic braking within 
the first core ineffective. As a result, a centrifugal disk can indeed 
form around the second core (the protostar).

The only other related MHD models that resolve the stellar core are the 
3D non-ideal MHD simulations by \citet{MachidaEtAl2006,MachidaEtAl2007} 
and \citet{MachidaMatsumoto2011}, who also used a simplified treatment 
of Ohmic dissipation. In most of their published models, the evolution 
was halted at up to ten days after the second core was formed, due to 
time-step restrictions, and the formation of a centrifugal disk was not 
studied. Very recent work \citep{MachidaMatsumoto2011} follows disk 
formation in one MHD model, and its relation to our work is discussed in 
Section \ref{sec:Discussion}.

In this paper, we increase realism by employing a detailed chemical network 
including the effect of grains to calculate partial ionization and 
the resulting Ohmic \textit{and} ambipolar diffusivities. We use the method 
presented in \citet{KunzMouschovias2009}. Both Ohmic dissipation and ambipolar 
diffusion gain importance with the formation of the first core 
\citep[e.g.,][]{LiMcKee1996, ContopoulosEtAl1998,DeschMouschovias2001}. 
We follow the evolution of the core to stellar sizes, using a multi-fluid approach that includes important inelastic collisions between gas-phase species and grains. We also study the effect of different grain sizes. We model four gas-phase species and up to five different grain sizes with three charge states (singly positively- or negatively charged, and neutral) for each grain size.
We demonstrate that Ohmic dissipation and ambipolar diffusion can reduce the efficiency of magnetic braking in the inner regions enough to allow for the formation of a centrifugally-supported disk. There is dramatic magnetic flux loss, and a small disk can form in a very early phase of evolution.
We propose a disk formation scenario in which the disk remains $\lesssim 10~\mathrm{AU}$ during the Class 0 phase,
and subsequently 
expands significantly due to internal angular momentum transport mechanisms such as magnetorotational instability (MRI) or gravitational torques \citep[e.g.,][]{BalbusHawley1991,VorobyovBasu2007}.

This paper is structured the following way. In Section \ref{sec:method_DBK11} we formulate the problem and describe the method of solution. Section \ref{sec:magn_braking} discusses the implementation of magnetic braking, while Section \ref{sec:MHD} outlines the derivation of the induction equation. Section \ref{sec:Chemistry} details the chemical model used to calculate the abundances of each species and Section \ref{sec:Initial_Conditions} contains the description of our initial conditions. Readers mainly interested in our results may proceed directly to Section \ref{sec:Results}. The discussion is found in Section \ref{sec:Discussion}, and we summarize our findings in Section \ref{sec:Summary_DBK11}.

\section{Method}\label{sec:method_DBK11}

We solve the equations of non-ideal MHD for a rotating, axisymmetric thin disk made 
up of partially-ionized gas \citep{BasuMouschovias1994}. In the thin-disk approximation, 
all quantities are assumed to be uniform over one scale height of the disk. It is applicable 
as long as the half-thickness of the disk remains small compared to the radius of the 
disk. \citet[][]{FiedlerMouschovias1993} showed that an initially cylindrical cloud 
quickly flattens parallel to the magnetic field before any significant collapse in the 
radial direction occurs, and a oblate cloud is formed that can be described by the 
thin-disk approximation \citep[see also][]{BasuMouschovias1994}. While this approximation 
breaks down at the interfaces of the first core and in the second core, it holds well in 
most of the cloud. It allows us to follow the evolution to very high densities and very 
small scales with realistic physics (non-ideal MHD and sophisticated chemical model), 
and does not require a lot of computation time.

We solve the following system of equations, derived by integrating the 
MHD equations over the $z$-direction and assuming axisymmetry: 
\begin{subequations}
\begin{align}
\frac{\partial \Sigma _{\mathrm{n}}}{\partial t} =
     &-\frac{1}{r} \frac{\partial}{\partial r}\left( r\Sigma _{\mathrm{n}}v_{r}\right),
     \label{eq:continuity} \\
\frac{\partial \left( \Sigma _{\mathrm{n}}v_{r}\right) }{\partial t}= 
     &-\frac{1}{r} \frac{\partial}{\partial r}\left( r\Sigma _{\mathrm{n}}v_{r}v_{r}\right) +
     f_{\mathrm{p}} + f_{\mathrm{g}} + f_{\mathrm{m}} + f_{\mathrm{r}},
     \label{eq:momentum} \\
\frac{\partial L}{\partial t}=
     &-\frac{1}{r}\frac{\partial}{\partial r} 
     \left(rLv_{r}\right)+\frac{1}{2\pi }rB_{z,\mathrm{eq}} B_{\varphi,Z},
     \label{eq:angmom} \\
\frac{\partial B_{z, \mathrm{eq}}}{\partial t} = 
		 &-\frac{1}{r} \frac{\partial}{\partial r} \left( r B_{z, \mathrm{eq}} v _{r}\right) \notag\\
		 &+\frac{1}{r} \frac{\partial}{\partial r}\left(r\eta_\mathrm{eff} 
		 \frac{\partial B_{z, \mathrm{eq}}}{\partial r}\right), 
		 \label{eq:induction}\\		 		 
P = &P(\varrho_{\mathrm{n}}).
\label{eq:barotropic_eqn}
\end{align} \label{eq:MHD_system}
\end{subequations}
Here, $\Sigma _{\mathrm{n}}$ is the mass column density of the neutrals, $v_{r}$ is the radial velocity, 
and $L\equiv\Sigma _{\mathrm{n}}\Omega r^2$ is the angular momentum per unit area. The mass volume 
density is denoted by $\varrho_{\mathrm{n}}$. The forces $f_{\mathrm{p}}$, $f_{\mathrm{g}}$, $f_{\mathrm{m}}$, 
and $f_{\mathrm{r}}$, as well as the magnetic field components ($B_{z,\mathrm{eq}}$, $B_{r,\mathrm{Z}}$, 
and $B_{\varphi ,\mathrm{Z}}$) are discussed below (see Eqs.(\ref{eq:forces}) as well as (\ref{eq:b_r_DBK11}) and (\ref{eq:b_phi_DBK11})). We do not include a separate continuity equation for the grains since no significant change of the dust-to-gas ratio occurs beyond a number density of $n\approx 10^{6}~\mathrm{cm}^{-3}$ \citep{CiolekMouschovias1996, KunzMouschovias2010}. Due to the very low ionization fraction in the modeled core, we neglect the inertia of all species but the neutral molecular gas.

We include the effect of both ambipolar diffusion \citep[e.g.,][]{MestelSpitzer1956,Mouschovias1979} \textit{and} Ohmic dissipation \citep[e.g.,][]{NakanoEtAl2002} in an effective resistivity $\eta _{\mathrm{eff}}$, which we derive from microphysical considerations in Section \ref{sec:MHD}. Note that we do not assume the resistivity to be spatially constant by pulling it outside all spatial derivatives, which is different from \citet[][]{MachidaEtAl2007}. We ignore the contribution of the third non-ideal MHD effect --- the Hall effect \citep[e.g.,][]{Wardle2007} --- in this Paper, and leave that for future work.

For computational ease, we use the barotropic relation Eq.(\ref{eq:barotropic_eqn}) instead of a detailed energy equation. 

Following DB10, we parametrize the temperature-density relation of \citet{MasunagaInutsuka2000} in the following way:

\begin{align}
  T =\left\{
		\begin{array}{ll}
				10                                        & n \le n _{1},\\
				10\left( \frac{n}{n_{1}}\right)^{5/3}     & n _{1} < n \le n _{2},\\
				10\left( \frac{n_{2}}{n_{1}}\right)^{5/3}
				  \left( \frac{n}{n_{2}}\right)^{7/5}     & n _{2} < n \le n _{3},\\
				10\left( \frac{n_{2}}{n_{1}}\right)^{5/3}
				  \left( \frac{n_{3}}{n_{2}}\right)^{7/5}
					\left( \frac{n}{n_{3}}\right)^{1.1}     & n _{3} < n \le n _{4},\\
				10\left( \frac{n_{2}}{n_{1}}\right)^{5/3}
				  \left( \frac{n_{3}}{n_{2}}\right)^{7/5}
					\left( \frac{n_{4}}{n_{3}}\right)^{1.1}
					\left( \frac{n}{n_{4}}\right)^{5/3}     & n _{4} < n.
		\end{array}%
		\right.    	
   \label{eq:Masunaga_Inutsuka}
\end{align}%

In this expression, the temperature $T$ is given in units of K. The number densities\footnote[1]{Note that during hydrogen dissociation the mass per particle is reduced from $2.3~m_{\mathrm{H}}$ (which is calculated assuming $10\%$ He in number) to $1.3~m_{\mathrm{H}}$, and then finally to $0.6~m_{\mathrm{H}}$ after the gas is fully ionized; $m_{\mathrm{H}}$ is the mass of a hydrogen atom. 
} $n$ at which the powerlaws in the temperature-density relation change are 
$n _{1} = 3\times 10^{10}~\mathrm{cm^{-3}}$, $n _{2} = 10^{13}~\mathrm{cm^{-3}}$, $n _{3} = 3\times 10^{16}~\mathrm{cm^{-3}}$, and $n _{4} = 10^{21}~\mathrm{cm^{-3}}$.

We transform this temperature-density relation into a pressure-density relation using the ideal gas law 
\begin{equation}
   P=nk_{\mathrm{B}}T.
   \label{eq:ideal_gas_law}
\end{equation}%
Here, $P$ is the thermal pressure and $k_{\mathrm{B}}$ is Boltzmann's constant. We calculate the thermal midplane pressure of the neutrals in our thin disk self-consistently, including the effects of the weight of the gas column, external pressure, magnetic pressure, and the effect of a central star (once present, with mass $M_{\bigstar }$):
\begin{equation}
    P_{\mathrm{n}}=\frac{\pi }{2}G\Sigma _{\mathrm{n}}^{2}+P_{\mathrm{ext}}+%
                   \frac{B_{r,Z}^{2}}{8\pi }+\frac{B_{\varphi,Z}^{2}}{8\pi } + W_{\bigstar }.
    \label{eq:neutral_pressure}
\end{equation}%
Here, $W_{\bigstar }$ is the extra vertical squeezing due to the star's gravity integrated over the disk's finite thickness. The external pressure is fixed at 
$P_{\mathrm{ext}} = 0.1~\pi G \Sigma _{\mathrm{0}} ^{2}/2$, where $G$ is the gravitational constant and $\Sigma _{\mathrm{0}}$ is the initial central neutral column density. We set $W_{\bigstar }=0$, and interpolate $\varrho_{\mathrm{n}}$ from the pressure-density relation found earlier.
If a central object is present (after the formation of the second core, and after introduction of the sink cell), we then insert the so-determined approximation to $\varrho_{\mathrm{n}}$ and the disk's half-thickness $Z=\Sigma_{\mathrm{n}}/2 \varrho_{\mathrm{n}}$ into 
\begin{equation}
		W_{\bigstar } = 2~GM_{\bigstar }\varrho_{\mathrm{n}} \int_{0}^{Z}\frac{z\mathrm{d}z}{\left(r^{2}+z^{2}\right) ^{3/2}},
		\label{eq:weight_pointmass}
\end{equation}
where $z$ is the vertical coordinate, and iterate this procedure until $\varrho_{\mathrm{n}}$ converges. The result also allows us to find the midplane temperature of our disk using Eq.(\ref{eq:ideal_gas_law}). It is very similar to that found in an axisymmetric $\left(r,z\right)$ simulation with explicit radiative transfer added \citep{KunzMouschovias2010}. Note that the extra squeezing from the protostar also flattens the disk further, helping to justify the thin-disk approximation.

Once the the second core forms, at a central number density of $\approx 10^{20}~\mathrm{cm}^{-3}$, we introduce a sink 
cell of size $\approx 3~\Rsun$ (which is only slightly larger than the second core). We remap the computational domain 
to a new grid with the sink as the innermost cell. The first neighboring cell outside the sink has an extent of only 
$\approx 10^{-3}~\mathrm{AU}\approx 0.2~\Rsun$, and each subsequent cell increases in size by a constant factor. We 
distribute matter within the sink cell so that much of it goes into a central point mass, but with enough remaining in 
the cell so that its column density remains equal to that in the first neighboring cell (because the gravitational solver
requires non-zero mass in the first cell). 
We also impose no-outflow boundary conditions on the sink cell. We modify the gravitational 
potential by adding the contribution of the central point mass (with mass $M_{\bigstar }$).

The forces per unit area appearing in the momentum equation Eq.(\ref{eq:momentum}) are given by:
\begin{subequations}
\begin{align}
f_{\mathrm{p}} &= -\frac{\partial}{\partial r} \left[ Z \pi G\Sigma _{\mathrm{n}}^{2}\right],
     \label{eq:pressure_force_DBK11} \\
f_{\mathrm{g}} &= \Sigma_{\mathrm{n}}g_{r},
     \label{eq:gravitational_force_DBK11} \\
f_{\mathrm{m}} &= \frac{B_{z,\mathrm{eq}}}{2\pi }\left( B_{r,Z}-
     Z\frac{\partial B_{z,\mathrm{eq}}}{\partial r}\right) \notag\\
		           &+ \frac{1}{4 \pi}\frac{\partial Z}{\partial r}\left(B_{r,Z}^{2}+B_{\varphi,Z}^{2}\right), 
     \label{eq:magnetic_force_DBK11} \\
f_{\mathrm{r}} &= \frac{L^{2}}{\Sigma _{\mathrm{n}}r^{3}}.  
     \label{eq:centrifugal_force_DBK11}
\end{align}
\label{eq:forces}
\end{subequations}
These expressions are derived in \citet[][]{CiolekMouschovias1993}. Equation (\ref{eq:pressure_force_DBK11}) is the thermal force in the thin-disk formulation due to pressure gradients, while Eq.(\ref{eq:gravitational_force_DBK11}) describes the radial gravitational force. Equation (\ref{eq:magnetic_force_DBK11}) represents the magnetic force due to azimuthal and poloidal components of the magnetic field at the top and bottom surfaces of the cloud, as well as from variations in scale height. This expression is derived in \citet{CiolekBasu2006}. Finally, Eq.(\ref{eq:centrifugal_force_DBK11}) is the centrifugal force in an axisymmetric thin-disk geometry, and couples the force equation with the equation for the evolution of the cloud's angular momentum, Eq.(\ref{eq:angmom}). 
Note that the effects of external pressure, magnetic pressure, and squeezing due to the protostar enter the pressure force through the half-thickness $Z$, and not explicitly in Eq.(\ref{eq:pressure_force_DBK11}).

We calculate the gravitational acceleration $g_{r}$ with the integral method for infinitesimally-thin disks employed in \citet{CiolekMouschovias1993} and \citet{MortonEtAl1994}. This produces a diverging field at the disk edge, which we avoid by adding in the gravitational effect of a virtual column density profile $\propto r^{-1}$ extending to infinity.
We also correct for the finite thickness of the flattened core so as not to overestimate the field strength \citep[see][]{BasuMouschovias1994}. 

The magnetic field points solely in the vertical direction inside the disk, and possesses radial and azimuthal components ($B_{r,Z}$ and $B_{\varphi ,Z}$) at the disk surfaces (indicated by subscript $Z$) and beyond. $B_{r,Z}$ is determined from a potential field assuming force-free and current-free conditions in the external medium, using the same integral kernel $\mathcal{M}\left( r,r^{\prime }\right)$ as for the gravitational field. We calculate $B_{\varphi,Z}$ and implement magnetic braking as in \citet{BasuMouschovias1994} (for details see Section \ref{sec:magn_braking}).

The remaining constituting equations are:
\begin{subequations}
\begin{align}
    B_{r,Z}\left( r \right)& =-\int_{0}^{\infty }\mathrm{d}r^{\prime }r^{\prime }\left[ B_{z,\mathrm{eq}}\left( r^{\prime } \right)-B_{\mathrm{ref}}\right] 
              \mathcal{M}\left( r,r^{\prime }\right), 
    \label{eq:b_r_DBK11}\\
    B_{\varphi ,Z}\left( r \right)
		     & =-2\frac{\sqrt{4\pi \varrho _{\mathrm{ext}}}}{B_{\mathrm{ref}}}
                     \frac{\Phi \left( r \right) }{r}\left[ \Omega \left( r \right) -\Omega _{\mathrm{ref}}\right], 
    \label{eq:b_phi_DBK11}\\
    \Phi\left( r \right) & =\int_{0}^{r}\mathrm{d}r^{\prime }r^{\prime }B_{z,\mathrm{eq}}\left( r^{\prime } \right), 
    \label{eq:magnetic_flux_DBK11}\\    
    g_{r}\left(r\right)
		     & =2\pi G\int_{0}^{\infty }\mathrm{d}r^{\prime }r^{\prime }\Sigma _{\mathrm{n}}
            \left( r^{\prime }\right) \mathcal{M}\left( r,r^{\prime }\right), 
    \label{eq:grav_potential_DBK11}\\
    \mathcal{M}\left( r,r^{\prime }\right) & =\frac{2}{\pi }\frac{\mathrm{d}}{\mathrm{d}r}\frac{1}{r_{>}}%
                                              K\left( \frac{r_{<}}{r_{>}}\right),
    \label{eq:integral_kernel}\\     
    Z\left( r \right)& =\frac{\Sigma _{\mathrm{n}}\left( r \right)}{2\varrho _{\mathrm{n}}\left( r \right)}.
    \label{eq:half_thickness_DBK11}
\end{align}
\end{subequations}
Here, $B_{\mathrm{ref}}$ is the assumed uniform background magnetic field of the surrounding medium with density $\varrho _{\mathrm{ext}}$, $\Phi$ is the magnetic flux, and $\Omega _{\mathrm{ref}}$ is the background rotation rate (see also Section \ref{sec:Initial_Conditions}). Lastly, $K$ is the Complete Elliptic Integral of the First Kind, and the symbols $r_{<}$ and $r_{>}$ denote the smaller and larger of $r$ and $r^{\prime }$, respectively.

We solve equations Eqs.(\ref{eq:continuity})-(\ref{eq:barotropic_eqn}) using the method of lines (e.g., Schiesser 1991) together with a finite volume approach. Hereby, the spatial part of the equations are initially discretized, and the resultant set of time-dependent ordinary differential equations (ODEs) are solved with LSODE \citep[][]{RadhakrishnanHindmarsh1993}. This implicit ODE solver uses the adaptive backward-difference method of \citet[][]{Gear1971}. The smallest cell size is initially $10^{-2}~\mathrm{AU}$ and becomes $0.02~\Rsun$ at the highest refinement. The advection step is done using a second-order van-Leer TVD advection scheme \citep{vanLeer1977}, and we calculate all derivatives to second-order accuracy on our nonuniform grid \citep[see][]{CiolekMouschovias1993}. We employ the method of \citet{NormanWilsonBarton1980} to advect angular momentum, which avoids angular momentum diffusion and associated spurious ring formation. 

Our computational grid has $256$ radial cells in logarithmic spacing. Test runs with $1024$ radial cells did not show any significant quantitative deviations. The grid is refined to a higher resolution (each step by a factor of $3$) whenever the column density increases by a factor of $50$. This allows us to satisfy the Truelove criterion \citep{TrueloveEtAl1997}, in that the Jeans length \citep{Jeans1902,Jeans1928} is resolved at all times by a minimum of $10$ cells.

Our boundary conditions are as follows. Besides the axial symmetry, we have reflection symmetry at the midplane and at the origin. Finally, at the outer radius, we have constant-volume boundary conditions.
For purposes of calculating derivatives we assume the column density to go to zero at the outer radius, and the magnetic field to go to its constant external value $B_{\mathrm{ref}}$.
The external medium has a low density $\varrho _{\mathrm{ext}}$ and a pressure $P _{\mathrm{ext}}$. 
The assumed high temperature and low density in the external medium allow
us to use the force-free and current-free approximation for the 
magnetic field above and below the cloud, as described earlier. 

\section{Magnetic braking} \label{sec:magn_braking}

We calculate magnetic braking using the same technique presented in \cite{BasuMouschovias1994} and also used by \citet[][]{KrasnopolskyKoenigl2002}, namely an analytical approximation to steady-state Alfv\'{e}n wave transport from the disk to an external medium. Owing to numerical complexity, a calibration of this method with results of three-dimensional MHD wave propagation through a stratified compressible medium has not been done to date. 

In order to derive the equations for magnetic braking, we consider the transport of angular momentum by torsional Alfv\'{e}n waves along flux tubes from the disk to the surrounding tenuous medium that has the density $\varrho _{\mathrm{ext}}$ and rotates at the angular frequency $\Omega _{\mathrm{ref}}$. The Alfv\'{e}n speed in the external medium is%
\begin{equation}
    v_{\mathrm{A,ext}}
      = \frac{B_{\mathrm{ref}}}{\sqrt{4\pi \varrho _{\mathrm{ext}}}}. 
\label{eq:Alfven_speed}
\end{equation}%
If the timescale for the Alfv\'{e}n waves to reach the background medium far
away from the disk is much smaller than the dynamical evolution timescale of
the disk, then the rate of angular momentum flux per unit radian through an
annulus at radius $r_{\mathrm{ref}}$ of thickness $\mathrm{d}r_{\mathrm{ref}}$ is 
\begin{equation}
    \frac{\mathrm{d}J}{\mathrm{d}t} = -\varrho _{\mathrm{ext}}r_{\mathrm{ref}}^{2}
                    \left( \Omega-\Omega _{\mathrm{ref}}\right) 
                    v_{\mathrm{A,ext}}r_{\mathrm{ref}}\mathrm{d}r_{\mathrm{ref}},  
\label{eq:AngMom_flux}
\end{equation}%
where $J$ denotes angular momentum per radian. The minus sign implies that we assume $\left( \Omega-\Omega _{\mathrm{ref}}\right) >0$, 
and so angular momentum is lost from the disk. In this expression we do not take into account any azimuthal drift of the field lines with respect to the 
neutrals, and leave that for a future study.
The radius $r_{\mathrm{ref}}$, far above the disk where the magnetic field has reached its uniform background state, 
is threaded by the same field line as radius $r$ in the disk. Eq.(\ref%
{eq:AngMom_flux}) considers the angular momentum that a volume of
gas of mass far away from the disk can take on. Angular momentum flux is
calculated by multiplying the angular momentum density $\varrho \Omega r^{2}$ 
with the transport velocity $v_{\mathrm{A,ext}}$ (which is constant far 
from the disk). 

We perform transformations to express Eq.(\ref{eq:AngMom_flux}) in terms of
quantities at the disk's surface instead of the external medium. First, we replace the external density
by the Alfv\'{e}n speed in Eq.(\ref{eq:Alfven_speed}).
Another assumption is that magnetic flux $\Phi$, defined by Eq.(\ref{eq:magnetic_flux_DBK11}), 
is conserved above the disk (flux freezing). Then, we can equate the flux in equatorial plane of the disk
$\Phi \left( r\right)$ with its value far above the disk, where the magnetic field 
$B_{\mathrm{ref}}$ is uniform. The foot point $r$ in the disk maps to a radius
$r_{\mathrm{ref}}$ above, since the two are connected by a flux tube. We have $r_{\mathrm{ref}}>r$ 
because the field lines are extending (diverging) above the disk. This means
\begin{subequations}
\begin{align}
    \Phi\left( r\right) &=\frac{1}{2}B_{\mathrm{ref}}r_{\mathrm{ref}}^{2},\\ 
    \mathrm{d}\Phi &= B_{\mathrm{ref}} r_{\mathrm{ref}} \mathrm{d}r_{\mathrm{ref}} 
                   = B_{z,\mathrm{eq}}r\mathrm{d}r.
\end{align}%
\end{subequations}
Using these relations, we can rewrite Eq.(\ref{eq:AngMom_flux}) to yield the
angular momentum flux%
\begin{equation}
    \frac{\mathrm{d}J}{\mathrm{d}t}=-\frac{\Phi }{2\pi v_{\mathrm{A,ext}}}
                                    \left( \Omega -\Omega _{\mathrm{ref}}\right) 
                                    B_{z,\mathrm{eq}}r\mathrm{d}r. 
\end{equation}%
Finally, taking into account the flux in both directions, above and below the disk (yielding a factor of 2),
and switching to angular momentum per unit area $L\equiv \Sigma_{\mathrm{n}} \Omega r^{2}=J\left/r\mathrm{d}r\right.$, we arrive at an expression for $N_{\mathrm{cl}}$, the torque acting on the cloud%
\begin{equation}
    \frac{\mathrm{d}L}{\mathrm{d}t}\equiv N_{\mathrm{cl}}=-\frac{\Phi }{\pi v_{\mathrm{A,ext}}}%
    \left( \Omega -\Omega _{\mathrm{ref}}\right) B_{z,\mathrm{eq}}. 
		\label{eq:MB_torque}
\end{equation}%
At the same time, the stress-energy tensor yields the change in angular
momentum to be equal to $rB_{z,\mathrm{eq}}B_{\varphi ,Z}/2\pi ,$ which
allows us to calculate the $\varphi $-component of the magnetic field at the
upper surface of the disk as%
\begin{align}
B_{\varphi ,Z} &=-\frac{2\Phi }{v_{\mathrm{A,ext}}}\frac{\left( \Omega
                 -\Omega _{\mathrm{ref}}\right) }{r}, \notag\\
               &=-2\frac{\sqrt{4\pi \varrho _{\mathrm{ext}}}}{B_{\mathrm{ref}}}
                 \frac{\Phi}{r}\left( \Omega -\Omega _{\mathrm{ref}}\right).
								\label{eq:B_phi}
\end{align}
Together, Eqs.(\ref{eq:MB_torque}) and (\ref{eq:B_phi}) yield the second term on the right hand side of the equation of angular momentum, Eq.(\ref{eq:angmom}). In the absence of magnetic braking, angular momentum is conserved and this term vanishes.

\section{Non-ideal MHD treatment}\label{sec:MHD}

We outline the derivation of the induction equation with all non-ideal MHD effects: \textit{ambipolar diffusion}, \textit{Ohmic dissipation}, and the \textit{Hall effect}. For brevity and clarity, we only consider a single grain size and omit inelastic collisions. However, we include their effect in our calculations, and refer the interested reader to the detailed exposition in Appendix B.1 in \citet{KunzMouschovias2009} where all terms are included. 

We start with Faraday's law in cgs units, transformed to the frame of the neutrals:
\begin{equation}
	  \frac{\partial \mathbf{B}}{\partial t} = c\mathbf{\nabla} \times \left( \frac{\mathbf{v}_{\mathrm{n}}}{c} \times \mathbf{B}
	  -\mathbf{E}_\mathrm{n}\right),
\label{eq:induction_equation_reference_neutral}
\end{equation}
where $\mathbf{B}$ is the magnetic field, $\mathbf{E}_{\mathrm{n}}$ is the electric field 
in the reference frame of the neutrals, and $c$ is the speed of light. In ideal MHD, 
$\mathbf{E}_\mathrm{n}\equiv0$ by definition, as the conductivity is infinite and so all local electric fields (in the neutral frame) are shorted by currents immediately. This is not the case in non-ideal MHD, where $\mathbf{E}_\mathrm{n} \neq 0$. We therefore seek an expression for $\mathbf{E}_\mathrm{n}$ in the general case.

We take the force equations for all charged species (denoted by subscript $s$, where in our discussion $s=\left\{\mathrm{i},\mathrm{e},\mathrm{g}^{+},\mathrm{g}^{-}\right\}$, but others would be possible), assuming force balance between the Lorentz force and collisions with neutrals. Inertial forces and collisions with other charged particles are neglected, as we are working under the assumption of a weakly-ionized plasma (with an ionization fraction $\chi \approx 10^{-8}$). Then we have
\begin{equation}
    0 = n_{s}q_{s} \left( \mathbf{E} + \frac{\mathbf{v}_{s}}{c} \times \mathbf{B}\right) - \frac{\varrho _{s}}{\tau _{s\mathrm{n}}}\left( 
    \mathbf{v}_{s}-\mathbf{v}_{\mathrm{n}}\right).
\label{eq:lorentz_force_eqn}
\end{equation}
The time scale for collisions between species $s$ and neutrals is given by $\tau _{s\mathrm{n}}$. The mass density of the charged species is $\varrho_{s}$, while $q_{s}$ is their charge number. Note that the latter carries a minus sign if the charge is negative (e.g., for electrons).

Introducing the \textit{drift velocity} $\mathbf{w}_{s} = \left(\mathbf{v}_{s} - \mathbf{v}_{\mathrm{n}}\right)$, we can transform Eq.(\ref{eq:lorentz_force_eqn}) to the reference frame of the neutrals:  
\begin{equation}
    0 = \omega _{s} \tau _{s\mathrm{n}} \left( \frac{c}{B}\mathbf{E}_{\mathrm{n}} + \mathbf{w}_{s} 
        \times \mathbf{b}\right) - \mathbf{w}_{s},
\label{eq:drift_velocity}
\end{equation}
where $\mathbf{b}\equiv \mathbf{B}/B$ is the unit vector in the direction of the magnetic field, and $B \equiv \left|\mathbf{B}\right|$ is the magnetic field strength. Also, we have introduced the \textit{cyclotron frequency} (in cgs units)
\begin{equation}
    \omega _{s} \equiv q_{s}B/m_{s}c
\label{eq:cyclotron_frequency}
\end{equation}
where $m_{s}$ is the mass of the charged particle, and note that $\varrho _{s}\equiv m_{s}n_{s}$ where $n_{s}$ is the number density of the charged species $s$.

We eliminate the term $\propto \left(\mathbf{w}_{s} \times \mathbf{b}\right)$ by inserting Eq.(\ref{eq:drift_velocity})$\times \mathbf{b}$ back into Eq.(\ref{eq:drift_velocity}). This yields
\begin{equation}
    \mathbf{w}_{s} = \omega _{s} \tau _{s\mathrm{n}} 
        \left( \frac{c}{B}\mathbf{E}_{\mathrm{n}} 
      + \omega _{s}\tau _{s\mathrm{n}}\frac{c}{B}\mathbf{E}_{\mathrm{n}}\times\mathbf{b} 
      - \omega _{s} \tau _{s\mathrm{n}}\mathbf{w}_{s,\perp}\right),
\label{eq:drift_velocity_complicated}
\end{equation}
which provides the drift velocity in terms of the electric field in the frame of the neutrals.

First, we look at the component parallel to $\mathbf{b}$:
\begin{equation}
    \mathbf{w}_{s,\parallel} =  \omega _{s}\tau _{s\mathrm{n}}
                              \frac{c}{B}\mathbf{E}_{\mathrm{n,\parallel}}  
\label{eq:drift_velocity_parl}																
\end{equation}
We write the \textit{electric current density} using the overall charge neutrality $\sum_{s}{n_{s}q_{s}} \equiv 0$:
\begin{align}
    \mathbf{j} = \sum_{s}{n_{s} q_{s}\mathbf{v}_{s}} = \sum_{s}{n_{s}q_{s}\mathbf{w}_{s}}.
\label{eq:currency_density}
\end{align}
and arrive at a generalized Ohm's law for the parallel component of $\mathbf{j}$
\begin{align}
    \mathbf{j}_{\parallel} &= \sum_{s}{n_{s} q_{s} \omega _{s}\tau _{s\mathrm{n}}\frac{ c }{B} 
                              \mathbf{E}_{\mathrm{n,\parallel}}} = \sum_{s}{\sigma _{s} 
                              \mathbf{E}_{\mathrm{n,\parallel}}}.
\label{eq:Ohms_law_parl}
\end{align}
Here,
\begin{equation}
    \sigma _{s} \equiv n_{s} q_{s}^{2} \tau _{s\mathrm{n}}\left/m _{s}\right.
\label{eq:conductivity_parl}
\end{equation}
is the conductivity due to species $s$. Lastly, we define $\sigma_{\parallel} = \sum_{s}{\sigma _{s}}$ as well as $\eta _{\parallel} = 1/\sigma_{\parallel}$ and invert Eq.(\ref{eq:Ohms_law_parl}) to get
\begin{equation}
    \mathbf{E}_{\mathrm{n,\parallel}} = \frac{1}{\sigma_{\parallel}}\mathbf{j}_{\parallel} = \eta _{\parallel} 
                                        \mathbf{j}_{\parallel}.
\end{equation} 

Similarly, we seek a relation between the perpendicular components of electric current 
density and electric field. Again, we start from Eq.(\ref{eq:drift_velocity_complicated}), 
but this time look at the perpendicular component:
\begin{align}    
    \mathbf{w}_{s,\perp}  &= \left(\frac{\omega _{s}\tau _{s\mathrm{n}}}{1+ \omega _{s}^{2} 
                             \tau _{s\mathrm{n}}^{2}} \frac{c}{B} \right) 
                             \left[\mathbf{E}_{\mathrm{n},\perp} + \omega _{s}\tau _{s\mathrm{n}}
                             \mathbf{E}_{\mathrm{n}}\times\mathbf{b} \right].
\end{align}
Inserting this into Eq.(\ref{eq:currency_density}), we find
\begin{align}
    \mathbf{j}_{\perp}  &= \sum_{s}{\left(\frac{\sigma_{s}}{1+\omega _{s}^{2}\tau _{s\mathrm{n}}^{2}} 
                           \right)} \mathbf{E}_{\mathrm{n},\perp}\notag\\
                        &+ \sum_{s}{\left( \frac{\sigma _{s}\omega _{s}\tau _{s\mathrm{n}}}
                             {1+ \omega _{s}^{2}\tau _{s\mathrm{n}}^{2}} \right)}  
                             \mathbf{E}_{\mathrm{n}}\times\mathbf{b}.
\end{align}
Defining 
\begin{equation}
     \sigma _{\perp} =\sum_{s}{\left(\frac{\sigma_{s}}{1+ 
                             \omega _{s}^{2}\tau _{s\mathrm{n}}^{2}} \right) } \quad \mathrm{and}\quad
     \sigma _{\mathrm{H}} = -\sum_{s}{\left( \frac{\sigma _{s}\omega _{s}\tau _{s\mathrm{n}}}
                             {1+ \omega _{s}^{2}\tau _{s\mathrm{n}}^{2}} \right)},  \notag
\end{equation}                             
Ohm's law for the perpendicular component is
\begin{equation}
    \mathbf{j}_{\perp} = \sigma _{\perp}\mathbf{E}_{\mathrm{n},\perp} - 
                         \sigma _{\mathrm{H}}\mathbf{E}_{\mathrm{n}}\times\mathbf{b}.
\end{equation}

Combining both parallel and perpendicular components, we can write down a generalized Ohm's law. Note that the magnetic field introduces an asymmetry (and thus off-diagonal components) and makes a tensor expression necessary:
\begin{align}
    \mathbf{j} &= 
		    \begin{pmatrix} 
				    \sigma _{\perp} & -\sigma _{\mathrm{H}} & 0 \\ 
		        \sigma _{\mathrm{H}} & \sigma _{\perp} & 0 \\ 
						0 & 0 & \sigma _{\parallel}  
				\end{pmatrix} 
		\mathbf{E}_{\mathrm{n}}.
\label{eq:Ohms_law_general}
\end{align}
Finally, we invert the matrix to get an expression for the electric field in the reference frame of the neutrals, 
\begin{equation}
    \mathbf{E}_{\mathrm{n}} = 
		    \begin{pmatrix} 
			      \eta _{\perp} & \eta _{\mathrm{H}} & 0 \\ 
		        -\eta _{\mathrm{H}} & \eta_{\perp} & 0 \\ 
						0 & 0 & \eta_{\parallel}  
			  \end{pmatrix} 
		\mathbf{j},
\end{equation}
where
\begin{equation}
     \eta _{\perp} \equiv \frac{\sigma _{\perp}}{\sigma _{\perp}^{2}+\sigma _{\mathrm{H}}^{2}}, \quad\quad  
     \eta _{\mathrm{H}}    \equiv \frac{\sigma _{\mathrm{H}}}{\sigma _{\perp}^{2}+\sigma _{\mathrm{H}}^{2}}, \quad \mathrm{and}\quad 
     \eta _{\parallel} = \frac{1}{\sigma _{\parallel}}.  \notag
\end{equation}

The purely vertical magnetic field in our thin disk is generated by an azimuthal current only. This means that there is no component of the electric current density parallel to the field (and neither in radial direction). Hence we are only interested in the perpendicular component of the electric field, and for convenience quote its expression:
\begin{align}
    \mathbf{E}_{\mathrm{n, \perp}} &= \eta _{\perp}\mathbf{j}_{\perp} + \eta _{\mathrm{H}}\mathbf{j}\times \mathbf{b},\notag\\
                                   &= \underbrace{\eta _{\parallel}\mathbf{j}_{\perp}}_{\mathrm{OD}} + 
                                      \underbrace{\left(\eta _{\perp}-\eta _{\parallel}\right)\mathbf{j}_{\perp}}_{\mathrm{AD}} + 
                                      \underbrace{\eta _{\mathrm{H}}\mathbf{j}\times \mathbf{b}}\label{eq:E_non_ideal_MHD}_{\mathrm{Hall}}.
\end{align}
Note that the quantity $\eta _{\perp} $ contains the effects of both ambipolar diffusion and Ohmic dissipation. In Eq.(\ref{eq:E_non_ideal_MHD}), the contributions of each of the three non-ideal MHD effects --- Ohmic dissipation (`OD'), ambipolar diffusion (`AD'), and the Hall effect (`Hall') --- are highlighted. We do not consider the Hall effect in the present work.

Summarizing the above derivation, we can write Eq.(\ref{eq:induction_equation_reference_neutral}) as
\begin{align}
    \frac{\partial B_{z,\mathrm{eq}}}{\partial t} &+ \frac{1}{r}\frac{\partial}{\partial r}\left( rB_{z,\mathrm{eq}}v_{r}\right)\notag \\ 
           &= -c \frac{1}{r}\frac{\partial}{\partial r}\left( rE_{n,\varphi}\right)
            = -c \frac{1}{r}\frac{\partial}{\partial r}\left( r\eta _{\perp}j_{\varphi}\right),\notag\\
           &= \frac{1}{r}\frac{\partial}{\partial r}\left( r\eta _{\mathrm{eff}} \frac{\partial B_{z,\mathrm{eq}}}{\partial r}\right),
\label{eq:non_ideal_MHD_induction_equation}
\end{align}
where the relation $\mathbf{j}=c/4\pi \mathbf{\nabla }\times\mathbf{B}$ has been used, and $\eta _{\mathrm{eff}}\equiv \eta _{\perp}c^{2}/4\pi $ was introduced. 

\section{Chemistry}\label{sec:Chemistry}

In this section, we present the chemical model used to calculate the ionization fraction, fractional abundances, 
and resistivities for seven species. In our models using an MRN grain-size distribution there are a total of 19 
species, but we omit the corresponding full chemical model (which can be found in \citealp{KunzMouschovias2009}) 
for sake of clarity. In the following, we consider neutral matter (one helium atom per five $\mathrm{H}_{2}$ molecules), 
atomic and molecular ions (such as $\mathrm{Mg}^{+}$ or $\mathrm{HCO}^{+}$), electrons, and grains (of uniform size, 
positively-charged, neutral, and negatively-charged). Multiply-charged grains are neglected, because the capture rate 
by a charged grain of a gas-phase particle with the same charge $q$ is reduced by a factor 
$\exp\left( - q^{2}/ak_{\mathrm{B}}T \right)$ \citep{Spitzer1941}, where $a$ is the particle's radius. A grain is 
thus far more likely neutralized by capturing a particle of opposite charge than elevated to higher charge by 
capturing a particle of the same charge. For example, \citet{NakanoEtAl2002} show that the abundance of 
doubly-charged grains is 5 orders of magnitude less than that of singly-charged ones. We also ignore multiply-charged 
ions and molecules.

\subsection{Ionization rate}\label{subsec:Ion_rate}

We consider four sources of ionization:
\begin{enumerate}
	\item UV ionization 
	\item cosmic ray ionization
  \item ionization due to radiation liberated in radioactive decay
  \item thermal ionization through collisions.
\end{enumerate}
UV ionization is only important where the visual extinction $\mathrm{A}_{\mathrm{V}} \lesssim 10$ \citep{CiolekMouschovias1995} or, equivalently, the $\mathrm{H}_{2}$ column density $N_{\mathrm{H}_{2}} \lesssim 2 \times 10^{22} ~\mathrm{cm}^{-2}$. The part of the cloud relevant to this work is much denser (in fact, everything within $10^4~\mathrm{AU}$ from the center is denser) and, to good approximation, UV ionization would not need to be considered. We still include its contribution in parametrized form, by adding $467.64~n_{\mathrm{H}_{2}}^{-2}~\mathrm{cm}^{-3}$ \citep[see][]{FiedlerMouschovias1992} to the electron and ion number densities. This has the effect to maintain an ionization fraction of $\approx 3\times 10^{-5}$ in the outermost envelope of the core, and keep it flux-frozen, but does not affect the dynamical evolution of higher-density regions.

In the higher-density regions (where $10^{4}~\mathrm{cm}^{-3}\lesssim n_{\mathrm{H}_{2}}\lesssim 10^{12}~\mathrm{cm}^{-3}$), the ionization is mainly due to cosmic rays. Their ionization rate is calculated by
\begin{equation}
  \zeta _{\mathrm{CR}} = \zeta _{0} \exp{\left[ - \Sigma _{\mathrm{H _{2}}}\right/96~\mathrm{g}~\mathrm{cm}^{-2}\left. \right]}
\label{eq:CR_ion_rate}
\end{equation}
\citep[][]{UmebayashiNakano1980}, where $\zeta _{0} = 5 \times 10^{-17}~\mathrm{s}^{-1}$ is the canonical unshielded cosmic-ray ionization rate \citep[][]{Spitzer1978}.

Beyond $n_{\mathrm{H}_{2}}\gtrsim 10^{12}~\mathrm{cm}^{-3}$, even cosmic rays are shielded and cannot penetrate deeply. Here, radioactivity, mainly due to 
$^{40}\mathrm{K}$, still provides a background level of ionization. The ionization rate is $\zeta _{\mathrm{40}} = 2.43 \times 10^{-23}~\mathrm{s}^{-1}$ 
\citep[e.g.,][]{KunzMouschovias2009}. Other radionuclides (such as $^{26}\mathrm{Al}$) are not considered, due to their low abundance, their short half-life 
time, their low ionization rate, or a combination thereof.

Finally, when the temperature reaches $\gtrsim 1000~\mathrm{K}$, collisions are energetic enough to cause thermal ionization of some atoms with low ionization potential (predominantly potassium). As the temperature rises further, collisions becomes the dominant source of ionization. We parametrize this as an additional source term in the ion equilibrium equation (see Section \ref{subsec:Abundances}) with the value \citep[see][]{PneumanMitchell1965}
\begin{align}
    \frac{\mathrm{d}n_{\mathrm{A}^{+}}}{\mathrm{d}t}
		      &= 2.9 \times 10^{-16}~\mathrm{cm}^{3}~\mathrm{s}^{-1}~n_{\mathrm{H}_{2}}n_{\mathrm{A}^{0}} 
             \left(T/1000~\mathrm{K}\right)^{1/2} \notag\\    
          &\times\exp\left(-5.03\times 10^{4}~\mathrm{K}/T\right), 
\label{eq:thermal_ion_rate}
\end{align}
where $n_{\mathrm{A}^{0}}$ is the fractional abundance of the neutral atomic particles. The fact that potassium only makes up $\approx 1/14$ of all metals \citep{Lequeux1975} has been considered in the coefficient in front.

Beyond $n_{\mathrm{H}_{2}}\gtrsim 10^{18}~\mathrm{cm}^{-3}$, at a temperature of $\gtrsim 1500~\mathrm{K}$, we assume the grains to be destroyed and the stored charges are released into the gas. The gas becomes highly ionized, and we thus revert back to flux freezing \citep{PneumanMitchell1965}. Note that the resistivity has already decreased significantly at this temperature as a consequence of thermal ionization.

\subsection{Fractional abundances}\label{subsec:Abundances}

In order to calculate the resistivity (see Section \ref{sec:MHD}), we calculate the fractional abundance of each species 
using a chemical equilibrium network. 

The molecular ion equation describes the production of molecular ions (such as $\mathrm{HCO}^{+}$) by radiative ionization, their destruction through charge-exchange reactions with neutral atoms and grains, as well as recombination reactions with electrons: 
\begin{align}
    \zeta n_{\mathrm{H}_{2}} 
    &= n_{\mathrm{m}^{+}}n_{\mathrm{A}^{0}}\beta 
     + n_{\mathrm{m}^{+}}n_{\mathrm{e}}\alpha_{\mathrm{dr}} \notag \\
    &+ n_{\mathrm{m}^{+}}n_{\mathrm{g}^{-}} \alpha_{\mathrm{m}^{+}\mathrm{g}^{-}}
     + n_{\mathrm{m}^{+}}n_{\mathrm{g}^{0}} \alpha_{\mathrm{m}^{+}\mathrm{g}^{0}}.
\label{eq:molecular_ions}
\end{align}
Here $\beta$ is the charge-exchange coefficient and $\alpha_{\mathrm{dr}}$ is the dissociative recombination rate (collisions with electrons). Their value and that of the other rate coefficients between species $s$ and $s\prime$, $\alpha_{s,s\prime}$, depend on temperature and grain properties, and are given in Appendix \ref{app:rate_coeffs}. The symbol $n_{s}$ refers to the number density of species $s$. Cosmic rays will ionize molecular hydrogen, forming $\mathrm{H}_{3}^{+}$ almost instantaneously; this in turn is highly reactive and will strip away an electron from any (non-$\mathrm{H}_{2}$) molecule it encounters. For instance, $\mathrm{CO}$ will be transformed to $\mathrm{HCO}^{+}$ and $\mathrm{H}_{2}$. This is why cosmic rays act on $\mathrm{H}_{2}$ in the first term of this equation.

Atomic ions (e.g., $\mathrm{Na}^{+}$) are produced by charge exchange reactions with neutral atoms, as well as thermal ionization, while they are destroyed by radiative recombinations with electrons, and by the collision with grains:
\begin{align}
    n_{\mathrm{m}^{+}}n_{\mathrm{A}^{0}}\beta &+ n_{\mathrm{A}^{0}}n_{\mathrm{H}_{2}}\alpha_{\mathrm{A}^{0}\mathrm{H}_{2}}
		= n_{\mathrm{A}^{+}}n_{\mathrm{e}}\alpha_{\mathrm{rr}} \notag \\
    & + n_{\mathrm{A}^{+}} n_{\mathrm{g}^{-}} 
       \alpha_{\mathrm{A}^{+}\mathrm{g}^{-}} 
    + n_{\mathrm{A}^{+}}n_{\mathrm{g}^{0}}\alpha_{\mathrm{A}^{+}\mathrm{g}^{0}},
\label{eq:atomic_ions}
\end{align}
where $\alpha_{\mathrm{rr}}$ is the radiative recombination rate, and the second term on the left-hand side represents thermal ionization (see Section \ref{subsec:Ion_rate}).

The equation for positively-charged grains balances the deposition of charge from atomic and molecular ions with the capture of electrons and neutralization with negatively-charged grains. The charge exchange between positively-charged and neutral grains is in steady-state, and so their contribution appears on both sides of the equation and cancels:
\begin{align}
       n_{\mathrm{m}^{+}}n_{\mathrm{g}^{0}}\alpha_{\mathrm{m}^{+}\mathrm{g}^{0}} 
    &+ n_{\mathrm{A}^{+}}n_{\mathrm{g}^{0}}\alpha_{\mathrm{A}^{+}\mathrm{g}^{0}} \notag\\ 
    &= n_{\mathrm{e}}n_{\mathrm{g}^{+}}\alpha_{\mathrm{e}\mathrm{g}^{+}}   
     + n_{\mathrm{g}^{+}}n_{\mathrm{g}^{-}}\alpha_{\mathrm{g}^{+}\mathrm{g}^{-}},
\label{eq:positive_grain}
\end{align}
Similarly, negatively-charged grains form by capture of an electron by a neutral grain, and are neutralized by charge exchange during collisions with molecular and atomic ions as well as positively-charged grains.
Again, the charge exchange between negatively-charged and neutral grains is in steady-state, and so does not appear:
\begin{align}
       n_{\mathrm{e}}n_{\mathrm{g}^{0}}\alpha_{\mathrm{e}\mathrm{g}^{0}}   
    &= n_{\mathrm{m}^{+}}n_{\mathrm{g}^{-}}\alpha_{\mathrm{m}^{+}\mathrm{g}^{-}}
     + n_{\mathrm{A}^{+}}n_{\mathrm{g}^{-}}\alpha_{\mathrm{A}^{+}\mathrm{g}^{-}}\notag\\ 
    &+ n_{\mathrm{g}^{+}}n_{\mathrm{g}^{-}}\alpha_{\mathrm{g}^{+}\mathrm{g}^{-}},
\label{eq:negative_grain}
\end{align}
Finally, to close the system, we include equations for the total number of atoms and grains
\begin{align}
    n_{\mathrm{A}^{0}} + n_{\mathrm{A}^{+}} &= n_{\mathrm{A}}\\
    n_{\mathrm{g}^{0}} + n_{\mathrm{g}^{+}} + n_{\mathrm{g}^{-}} &= n_{\mathrm{g}},     
\label{eq:total_particles}
\end{align}
as well as for overall charge neutrality
\begin{align}
    n_{\mathrm{m}^{+}} + n_{\mathrm{A}^{+}} + n_{\mathrm{g}^{+}} - n_{\mathrm{g}^{-}} - n_{\mathrm{e}} &= 0.
\label{eq:charge_neutrality}
\end{align}

Equations (\ref{eq:molecular_ions})-(\ref{eq:charge_neutrality}) form the non-linear system to be solved for the fractional abundances; we do this iteratively for each time step and each grid point using Newton's method. The linearized matrix equation that results involves the Jacobian and the update vector to the abundances, and is solved directly using a LU-decomposition package. We assume the mean mass of the molecular and atomic species to be $m_{\mathrm{m}^{+}}=29~m_{\mathrm{p}}$ and $m_{\mathrm{A}^{+}}=23.5~m_{\mathrm{p}}$, respectively, where $m_{\mathrm{p}}$ is the proton mass. Those values are the masses of $\mathrm{HCO}^{+}$ and the average of atomic magnesium and sodium, respectively, which are good representatives of the broader range of species present.

\section{Initial Conditions and Physical Units}\label{sec:Initial_Conditions}

We assume that our initial state was reached by core contraction preferentially along magnetic field lines \citep[e.g.,][]{FiedlerMouschovias1993} and rotational flattening, and prescribe initial profiles for column density and angular velocity given by
\begin{align}
	\Sigma \left( r \right)		= \frac{\Sigma _{0}}{\sqrt{1+\left(r/R\right)^{2}}},&&
	\Omega \left( r \right)		= \frac{2 \Omega _{\mathrm{c}}}{\sqrt{1+\left(r/R\right)^{2}}+1}.		
	\label{eq:initial_conds}
\end{align}%
Here, $R\approx 1,500~\mathrm{AU}$ approximately equals the Jeans length at the core's initial central 
density (see below). The column density profile is representative of the early stage of collapse 
\citep[e.g.,][]{Basu1997,DappBasu2009}, and the angular velocity profile reflects that the specific 
angular momentum of any parcel is proportional to the enclosed mass $m_{\mathrm{encl}}$.

We assume an initial profile for $B _{z, \mathrm{eq}}$ such that initially the normalized mass-to-flux ratio $\mu=2 \pi \sqrt{G}~\Sigma / B_{z, \mathrm{eq}}=2$ everywhere, which is the approximate starting point of runaway collapse \citep[e.g.,][]{BasuMouschovias1995b}, and also consistent with observed values in molecular cloud cores \citep{Crutcher1999}. The core is at rest at the beginning of our simulations, i.e., the radial velocity is zero. The initial state is not far from equilibrium, as the pressure gradient and magnetic and centrifugal forces add up to $\approx 82\%$ of the gravitational force. Our results do not depend strongly on the choice of initial state, as long as gravity remains dominant. 

The initial central column density is set to $\Sigma _{0} = 2 \times 10^{-2}~\mathrm{g}~\mathrm{cm}^{-2}$. The core 
has a temperature $T=10~\mathrm{K}$, and its total mass and radius are $28.5~\Msun$ and $0.6 ~\mathrm{pc}$, respectively. 
The initial central number density and magnetic field strength are $n _{\mathrm{c}} = 3.3\times 10^{4}~\mathrm{cm}^{-3}$ 
and $B_{z, \mathrm{eq}}\approx 200~\mu \mathrm{G}$, respectively. We choose the external density to be 
$n _{\mathrm{ext}} = 10^{3}~\mathrm{cm}^{-3}$, and the central angular velocity $\Omega _{\mathrm{c}}$ such that the 
cloud's edge rotates at a speed of $0.1~\mathrm{km}~\mathrm{s}^{-1}~\mathrm{pc}^{-1}$. 

Transient adjustments occur if the simulation is started from an initial non-equilibrium state that is at rest. The chemistry calculations are quite sensitive to fluctuations in density, which can cause problems. We therefore let the system evolve from an initial state with the above-mentioned profiles and characteristics, but initially without non-ideal MHD effects. Once the cloud has settled into a steady infall (at a central density of approximately $n _{\mathrm{c}} \approx 4\times 10^{6}~\mathrm{cm}^{-3}$) the full MHD equations Eqs.(\ref{eq:continuity})-(\ref{eq:barotropic_eqn}) are solved. The state at which we switch on the detailed treatment of chemistry corresponds very closely to the initial state employed by DB10, so that comparisons with their results are possible. The rotation rate by then has increased to $1~\mathrm{km}~\mathrm{s}^{-1}~\mathrm{pc}^{-1}$, consistent with observations of large molecular cloud cores \citep{GoodmanEtAl1993,CaselliEtAl2002}. In the plots in this paper, we show only the region of the core within $0.05 ~\mathrm{pc}$, again consistent with DB10. Note that the nature of the collapse is very dynamical and happens under flux freezing to a very good approximation between $n _{\mathrm{c}} \sim 10^{4}-10^{10}~\mathrm{cm}^{-3}$ \citep[see][]{KunzMouschovias2010}. 

We fix the dust-to-gas ratio at $1\%$, consistent with observations \citep[][]{Spitzer1978}, 
and keep it constant everywhere and at all times. This means that a different mean grain size will 
result in differing total numbers of grains available, by a factor of $\left(a_{\mathrm{gr}}/a_{\mathrm{gr,min}}\right)^{-3}$, with a total surface 
area $\propto \left(a_{\mathrm{gr}}/a_{\mathrm{gr,min}}\right)^{-1}$. 

\section{Results}\label{sec:Results}

\subsection{Collapse phase}\label{subsec:results_collapse}

We ran multiple models with different single grain sizes, and one with a standard ``MRN'' distribution (named after its authors: \citealp*[][]{MathisEtAl1977}). In this case, the number density of spherical dust grains with radii between $a$ and $a + \mathrm{d}a$ is
\begin{equation}
  \mathrm{d}n_{\mathrm{g,tot}}\left( a \right) = N_{\mathrm{MRN}} a^{-3.5} \mathrm{d}a .
\end{equation}
The distribution is truncated at a lower grain radius $a_{\mathrm{min}}=0.011~\mathrm{\mu m} $ and an upper grain 
radius $a_{\mathrm{max}}=0.26~\mathrm{\mu m}$. The coefficient $N_{\mathrm{MRN}}$ is proportional to the dust-to-gas 
mass ratio in the cloud. \citet[][]{KunzMouschovias2009} found that there is reasonable convergence for 5 grain size 
bins, so we choose that number in order to limit computational expense. More detail can be found in that paper (their Section 4.2). 
Our models are summarized in Table \ref{tab:table_models_DBK11}.

\ctable[
cap         =  Simulation model overview.,
caption     =  Simulation model overview.,
label       =  tab:table_models_DBK11,
mincapwidth =  0.6\hsize,
]{ccc}{
\tnote{All models above start from the same initial conditions, comparable to those described in DB10. For details, see Section \ref{sec:Initial_Conditions}.}
}{ \FL
 Model         &  $a_{\mathrm{gr}} /\mu m$  & Notes\ML
1    & 0.019 & --- \NN
2    & 0.038 & fiducial grain size \NN
3    & 0.075 & --- \NN
4    & 0.113 & --- \NN
5    & 0.150 & --- \NN
6    & 0.038 & no resistivity\NN
7    & 0.038 & no magnetic braking\NN
8    & ---   & OD alone (as in DB10)\NN
9    & MRN distribution & --- \LL
}

\begin{figure}[htp]%
    \includegraphics[width=0.9\hsize]{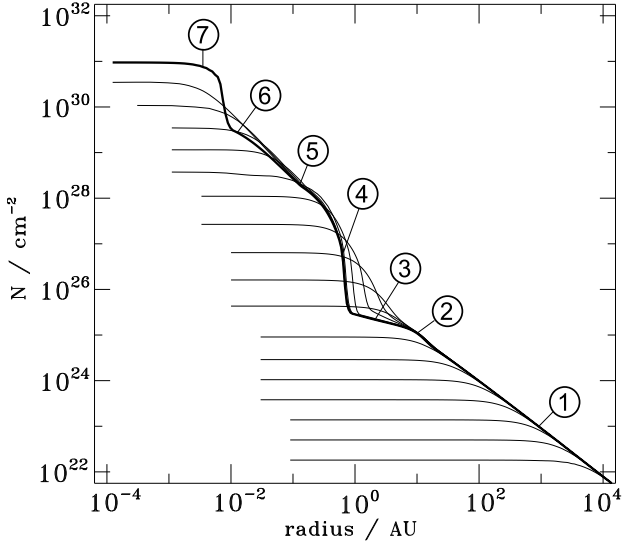}%
		\centering
    \caption[Column number density versus radius over time]
            {Column number density profile versus radius for Model 2. The thin lines (in ascending order) are plots 
						at the times listed in Table \ref{table:times}.
						Several features are identifiable via their associated breaks in the profile.
						(1) Prestellar infall profile with $N \propto r^{-1}$. (2) Magnetic wall at 
						$\approx 10~\mathrm{AU}$, where the bunched-up field lines
						decelerate material before it continues the infall. (3) Expansion wave profile with 
						$N \propto r^{-1/2}$ outside the first core. 
						(4) First core at $1~\mathrm{AU}$. (5) Infall profile onto the second core with $N \propto r^{-1}$.
						After the first core has reached $\approx 1,000~\mathrm{K}$, it starts to collapse, as
						H$_{2}$ is dissociated. (6) Expansion wave profile with 
						$N \propto r^{-1/2}$ outside the second core.  
						(7) Second core at $\approx 1~\Rsun$. 
						}
    \label{fig:sigma_time_DBK11}
\end{figure}

Figure \ref{fig:sigma_time_DBK11} shows the column number density profile versus radius at different times 
for Model 2 (with $a_{\mathrm{gr}}=0.038~\mathrm{\mu m}$). Several features are identifiable via 
their associated breaks in the profile. From the outside in, those are:
\begin{enumerate}
	\item Prestellar infall profile with $N \propto r^{-1}$. This corresponds to a volume number density profile of 
	      $n \propto r^{-2}$, typical for collapsing cores dominated by gravity (see, e.g., 
				\citealp[][]{Larson1969}; \citealp{DappBasu2009}). The vertical hydrostatic condition mandates that $n \propto N^{2}$.
	\item At $\approx 10~\mathrm{AU}$, a \textit{magnetic wall} \citep{LiMcKee1996,ContopoulosEtAl1998,TassisMouschovias2007b}. 
	      Here, the relatively well-coupled, bunched-up magnetic field expelled (relative to the neutrals) from the first core decelerates material 
				temporarily. At smaller radii, the infall continues. 
	\item Expansion wave profile with $N \propto r^{-1/2}$ outside the first core. 
The power law can be motivated analytically
				\citep[see][for the spherical case]{Shu1992}. Energy conservation requires the infall speed from 
				large distances towards a point mass (i.e., the first core) with mass $M$ to scale as $v_{r}\propto \sqrt{GM/r}\propto r^{-1/2}$. At the 
				same time the infall onto the point mass is essentially a steady-state process, and thus
				$\dot M \equiv 2 \pi r \Sigma v_{r} = \mathrm{const}$, close to the border of the first core. Together, those
				two relations imply $\Sigma \propto r^{-1/2}$. 
	\item First core at $1~\mathrm{AU}$. Here, the density is sufficiently high that the gravitational energy released 
	      in the collapse cannot escape as radiation anymore. The temperature rises, and thermal pressure gradient stabilizes the 
				object. The first core is nearly in hydrostatic equilibrium, and its radial and vertical extent are approximately equal.
	\item Infall profile onto the second core with $N \propto r^{-1}$. After the temperature in the first core has reached 
	      $\approx 1,000~\mathrm{K}$, hydrogen dissociates. This process provides a heat sink, and the temperature 
				no longer increases sufficiently for thermal pressure to balance gravity. The core starts to collapse again and flattens at the same time.
	\item Beginning expansion wave profile with $N \propto r^{-1/2}$ outside the second core, for the same reasons as outside
	      the first core. Once this rarefaction wave reaches the boundary of the first core, the material comprising the first core 
				will fall in to a region of stellar dimensions, unless it forms a centrifugal disk. 
	\item Second core at $\approx 1~\Rsun$. After hydrogen dissociation (and ionization) has concluded, the temperature rises again, 
        and a truly hydrostatic object, the YSO, is formed. It is also partly supported by electron degeneracy pressure  
				\citep{MasunagaInutsuka2000}. 
\end{enumerate}

\ctable[
cap         =  Output times and number density in the profile plots.,
caption     =  Output times\tmark ~and corresponding central number density in the profile plots.,
label       =  table:times,
mincapwidth =  1.0\hsize,
]{ccc}{
\tnote{Time is counted from when chemistry and non-ideal MHD effects are switched on (see Section \ref{sec:Initial_Conditions}).}
\tnote[b]{Lines count from bottom upwards.}
}{ \FL
 Line number\tmark[b]  &  time / $10^{3}~\mathrm{yr}$  & $n_{\mathrm{c}}$ / cm$^{-3}$\ML
1  & $0.0$            & $4.3 \times 10^{6}$ \NN 
2  & $14.338$      & $3.3 \times 10^{7}$ \NN
3  & $19.303$      & $2.5 \times 10^{8}$ \NN
4  & $21.063$      & $1.9 \times 10^{9}$ \NN
5  & $21.692$      & $1.4 \times 10^{10}$ \NN
6  & $21.944$      & $1.1 \times 10^{11}$ \NN
7  & $22.045$      & $8.6 \times 10^{11}$ \NN
8  & $22.085$      & $6.8 \times 10^{12}$ \NN
9  & $22.105$      & $4.8 \times 10^{13}$ \NN
10 & $22.116$     & $3.7 \times 10^{14}$ \NN
11 & $22.139$     & $2.8 \times 10^{15}$ \NN
12 & $22.1508$   & $2.1 \times 10^{16}$ \NN
13 & $22.15137$ & $1.6 \times 10^{17}$ \NN
14 & $22.15149$ & $1.2 \times 10^{18}$ \NN
15 & $22.15153$ & $9.3 \times 10^{18}$ \NN
16 & $22.15154$ & $7.0 \times 10^{19}$ \NN
17 & $22.15156$ & $1.6 \times 10^{20}$ \LL
}

\begin{figure}[htp]%
    \includegraphics[width=0.9\hsize]{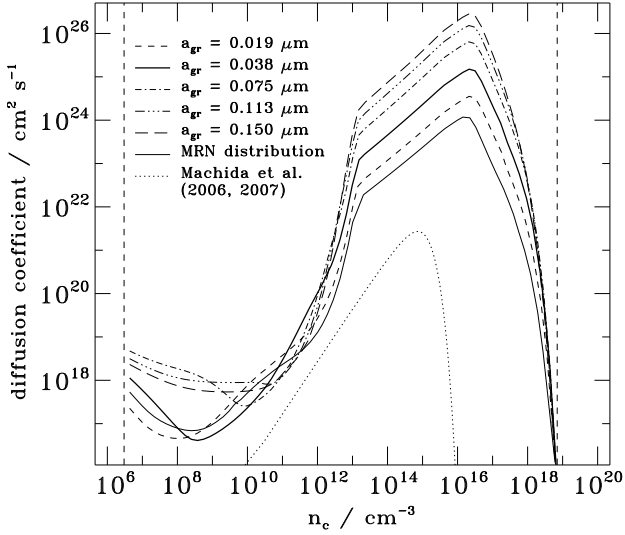}%
    \centering
		\caption[{Diffusion coefficient versus central density for different grain sizes}]
             {Central diffusion coefficient $\eta_{\rm eff}$ versus number density for different grain sizes, extracted from the dynamical model. 
						The vertical line at the left indicates the density at which the detailed chemistry 
						and non-ideal MHD treatment is switched on. Beyond 
						$n_{\mathrm{c}}\approx 10^{18}~\mathrm{cm}^{-3}$ the resistivity plummets, 
						after having already declined due to thermal ionization. This 
						is where we switch the chemistry calculations off again, and is denoted by the vertical line on the right. Due to grain 
						destruction, flux-freezing is restored there.
						}
    \label{fig:diffc_DBK11}
\end{figure}

Figure \ref{fig:diffc_DBK11} shows the evolution of the central diffusion coefficient $\eta_{\rm eff}$ versus number
density for the various models (different single grain sizes and the MRN 
distribution of grain sizes; see Table \ref{tab:table_models_DBK11}), as well as the parametrized resistivity used 
in \citet{MachidaEtAl2006} and DB10. At low densities, electrons and ions are 
the dominant (but not exclusive) contributors to the conductivity. Smaller grains are fairly well 
coupled to the magnetic field because of their smaller gyro-radii and smaller collisional 
cross sections, so their contribution to the effective resistivity is lower than that of larger grains. 
However, there is a competing effect: smaller grains have an increased capability to absorb gas-phase 
charge carriers because of their larger total surface area. For grains with radius $a_{\mathrm{gr}}=0.075~\mathrm{\mu m}$ 
(dot-dashed line) and larger, the trend reverses and the conductivity increases (the resistivity 
decreases) with larger grain radius, as expected from an increased ionization fraction. 

\begin{figure}[htp]%
    \includegraphics[width=0.9\hsize]{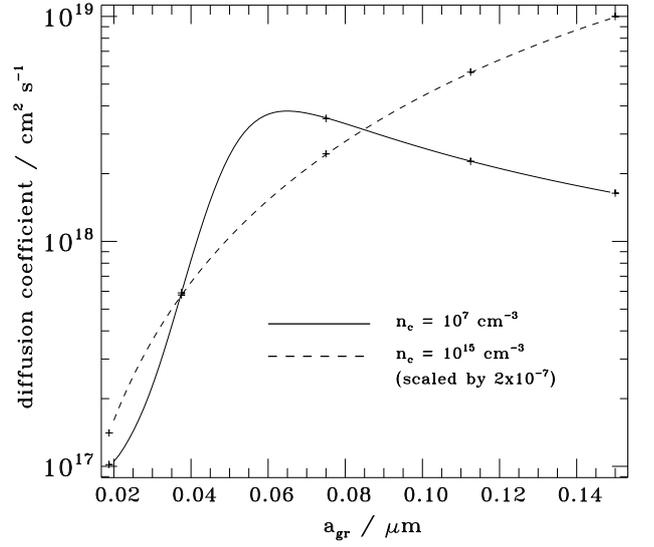}%
    \centering
		\caption[{Diffusion coefficient versus grain size for two values of number density}]
             {Diffusion coefficient versus grain size for two values of number density. 
              The diffusion coefficient behaves qualitatively differently at different densities. While
              at high densities, the diffusion coefficient increases monotonically with increasing 
              grain radius, it has a maximum for low densities at a grain size of 
              $a_{\mathrm{gr}}\approx 0.065~\mathrm{\mu m}$. Below this radius, the effect of better
              coupling dominates, while at larger grain sizes the decreased surface area of the larger
              grains leaves more free gas-phase charges, with associated higher conductivity, and 
              conversely decreased resistivity.
			 }
    \label{fig:diffc_agr_DBK11}
\end{figure}

This effect is demonstrated in Fig. \ref{fig:diffc_agr_DBK11} which shows the diffusion coefficient 
versus grain size at two densities ($n\approx 10^{7}~\mathrm{cm}^{-3}$ -- solid line; 
$n\approx 10^{15}~\mathrm{cm}^{-3}$ -- dashed line). The symbols indicate values for which 
we have complete runs, while the continuous lines are computed by calculating the effective resistivity 
with values for temperature, column and volume density, as well as magnetic field averaged over the values from the available complete runs. The magnetic field strength
enters the resistivity only though the gyro-frequency, and the estimated error
in the continuous lines is typically $<1\%$. 
The diffusion coefficient takes on a maximum 
at a grain size of $a_{\mathrm{gr}}\approx 0.065~\mathrm{\mu m}$ for low densities (solid line in Fig. 
\ref{fig:diffc_agr_DBK11}). As described above, for grains with smaller radius, the effect of better 
coupling dominates, while at larger grain sizes, the associated decrease in total surface area leaves more 
free gas-phase charges, and makes for a higher conductivity, and conversely a decreased resistivity.

At intermediate densities the resistivity in Fig. \ref{fig:diffc_DBK11} rises sharply, as a consequence 
of the grains soaking up gas-phase charges and decoupling from the magnetic field. 
Additionally, the conductivity drops because at high column density cosmic rays are shielded 
according to Eq.(\ref{eq:CR_ion_rate}). At $n_{\mathrm{c}} \approx 10^{13}~\mathrm{cm}^{-3}$ 
the only remaining source of ionization is radioactivity, which provides a floor ionization rate
(see Section \ref{subsec:Ion_rate}). At this stage, clearly distinguished by a break in the 
profile, the resistivity is almost exclusively due to grains, i.e., their collisions with 
neutrals. Inserting Eq.(\ref{eq:collision_times}) of Appendix \ref{app:collision_times} 
into Eq.(\ref{eq:conductivity_parl}), we see that conductivity scales as $a_{\mathrm{gr}}^{-2}$ in this 
phase. The resistivity is the inverse of the conductivity, and hence it \textit{increases} 
with larger grains as a consequence of their lower gyro-frequencies and greater collision 
cross sections. This happens even though there are fewer grains with increasing grain size, and 
they have a reduced total surface area. The dashed line in Fig. \ref{fig:diffc_agr_DBK11} shows this monotonic 
increase in diffusion coefficient with grain radius at a central density of 
$n_{\mathrm{c}} \approx 10^{15}~\mathrm{cm}^{-3}$. As the temperature 
approaches $\approx 1,000~\mathrm{K}$, collisions become energetic enough that thermal ionization
occurs, described by Eq.(\ref{eq:thermal_ion_rate}). The conductivity recovers, and hence the resistivity 
decreases again. Finally, during the second collapse, as temperatures of $\approx 1500~\mathrm{K}$ 
are reached, grains are destroyed and all locked-up charges are released into the gas phase. Electrons 
and ions flood the gas, the ionization fraction rises sharply, and flux-freezing is restored 
(see Section \ref{subsec:Ion_rate}). We switch off the chemistry calculations at this point, shown by the right vertical dashed line in Fig. \ref{fig:diffc_DBK11}). We point out that the parametrization for only Ohmic 
dissipation used in \citet[][]{MachidaEtAl2006, MachidaEtAl2007} and DB10 (for comparison shown in Fig. \ref{fig:diffc_DBK11} as the dotted line) yields values for the resistivity lower by at least 
a factor of $10$ everywhere.

In Model 9 (the MRN distribution), the diffusion coefficient
is dominated by the smallest grain size present, and behaves otherwise very similar to the simulations with a 
single (small) constant grain size.

Note that Fig. \ref{fig:diffc_DBK11} shows data extracted from the central cell of a dynamical simulation. The exact value of the 
diffusion coefficient not only depends on the grain size and the local number density of the neutrals, but also on the number densities of
the other species and the local values of magnetic field strength, temperature, and column density. 
As a result, the values shown in this plot are not necessarily representative of any other point in the cloud. 
It is therefore not helpful to parametrize the diffusion coefficient according to this plot and simply use an approximate 
interpolative expression as a replacement for a realistic calculation of the microphysics and ionization balance. 
The procedure outlined in Sections \ref{sec:MHD} and \ref{sec:Chemistry} is the simplest method of obtaining accurate values 
for the diffusion coefficient. 

\begin{figure}[htp]%
    \includegraphics[width=0.9\hsize]{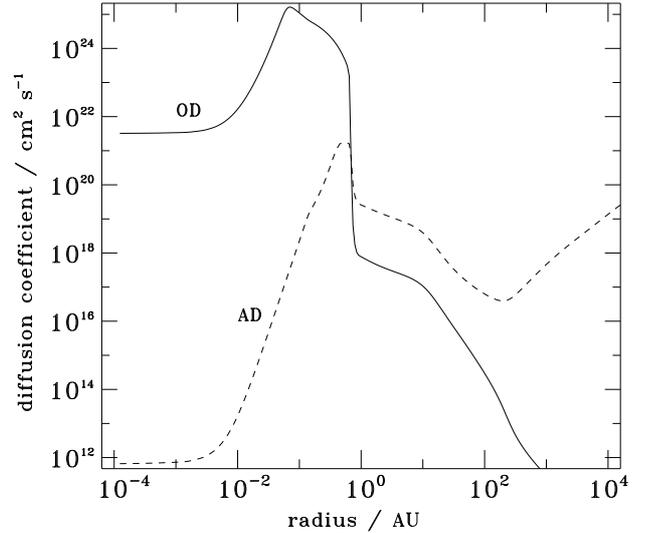}%
		\centering
    \caption[{Contributions to the effective resistivity}]
             {Contributions due to ambipolar diffusion (AD) and Ohmic dissipation (OD)
						  to the effective resistivity in Model 2 
							(grain size $a_{\mathrm{gr}}=0.038~\mathrm{\mu m}$) at the end of the run. OD only dominates
							AD beyond $n_{\mathrm{c}} \gtrsim 10^{12}~\mathrm{cm}^{-3}$, within 
                                                        the first core.
						 }
    \label{fig:eta_DBK11}
\end{figure}

Figure \ref{fig:eta_DBK11} shows the relative contribution of ambipolar diffusion (AD) and Ohmic dissipation (OD)
to the resistivity coefficient near the end of the run. 
The coefficient is determined mainly by AD everywhere outside the first core at $r \approx 1~\mathrm{AU}$. In fact, AD dominates OD up to a density of $n_{\mathrm{c}}\approx 5\times 10^{12}~\mathrm{cm}^{-3}$. The contribution of OD continues to increase sharply at higher densities and is the dominant non-ideal MHD effect within the first core. Its coefficient only decreases again in the vicinity of the second core at $\approx 0.1~\mathrm{AU}$, where thermal ionization drives up the conductivity. Note that the comparatively large AD diffusion coefficient outside $\approx 10^{2}~\mathrm{AU}$ does not cause a large flux loss since the dynamical time is still much smaller than the diffusion time associated with AD.

\begin{figure}[htp]%
    \includegraphics[width=0.9\hsize]{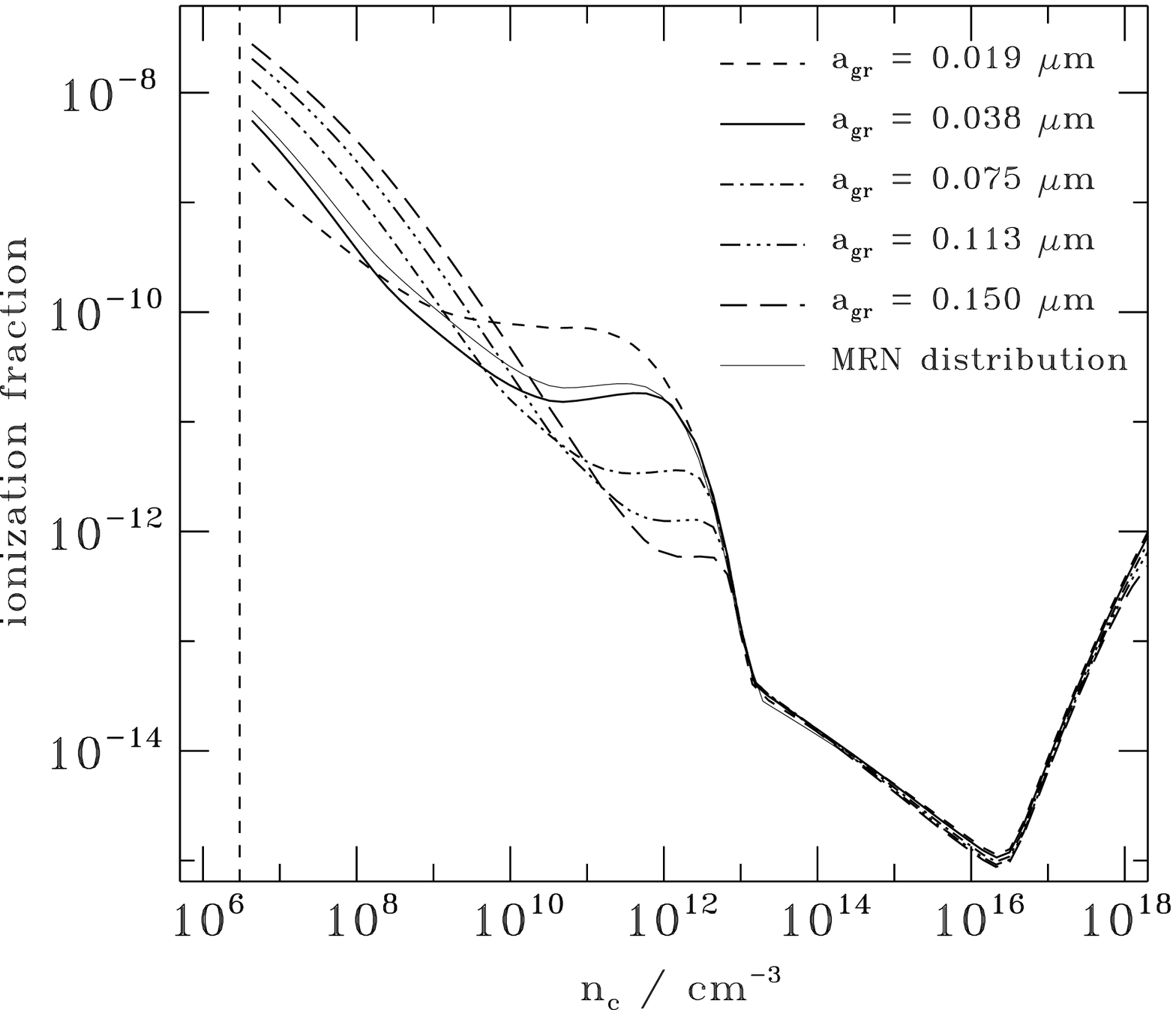}%
		\centering
    \caption[{Ionization fraction versus central density for different grain sizes}]
             {Total ionization fraction versus central density for different grain sizes.
						 The vertical line indicates the density at which the detailed chemistry 
						 and non-ideal MHD treatment is switched on.
						 }
    \label{fig:ionfrac_DBK11}
\end{figure}

\begin{figure*}[htp]
	\begin{minipage}[b]{0.46\hsize}
		\centering
		
		\includegraphics[width=1.0\hsize]{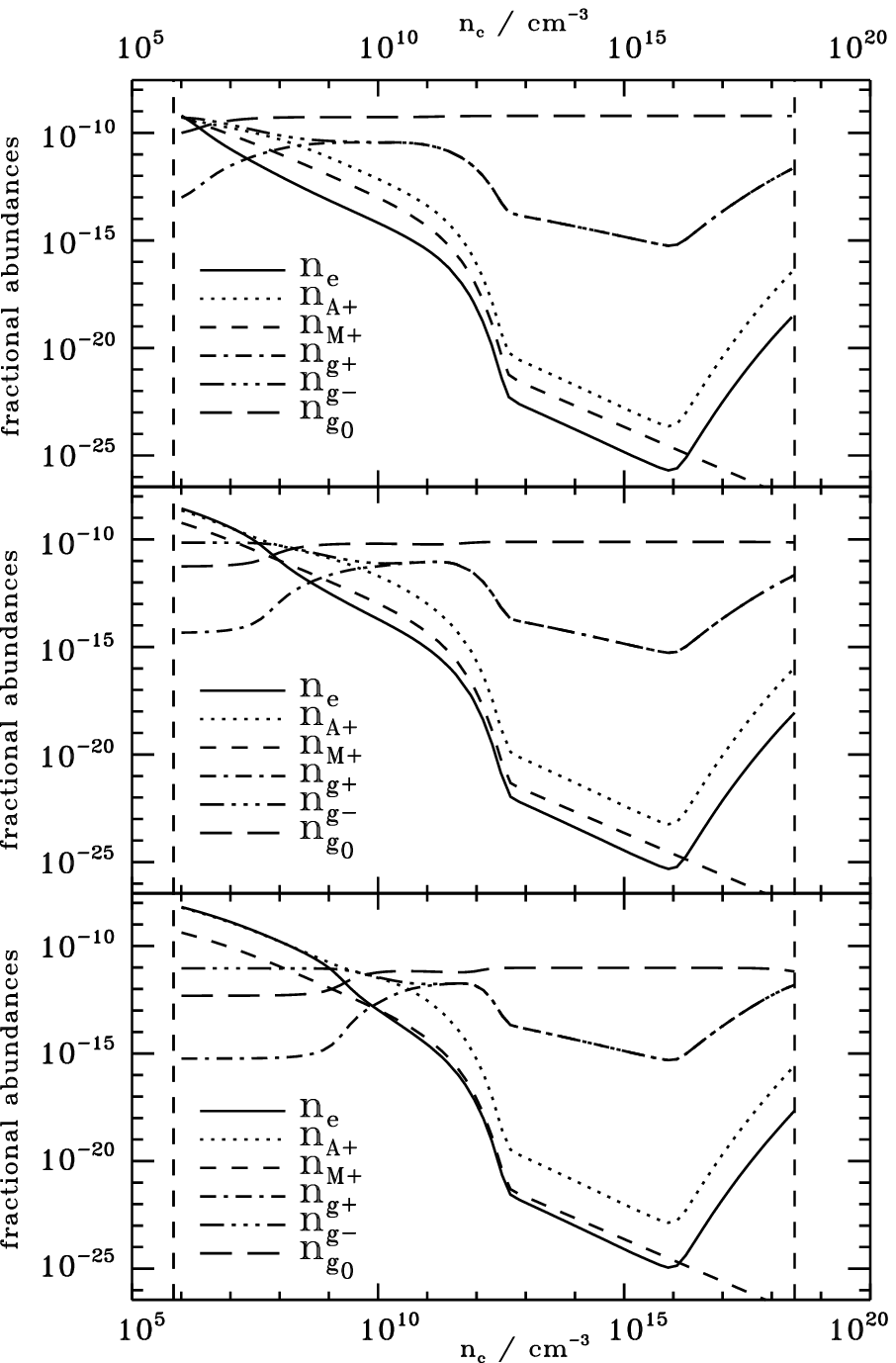}		
	\end{minipage}
	\hspace{0.5cm}
	\begin{minipage}[b]{0.46\hsize}
		\centering
		\includegraphics[width=1.0\hsize]{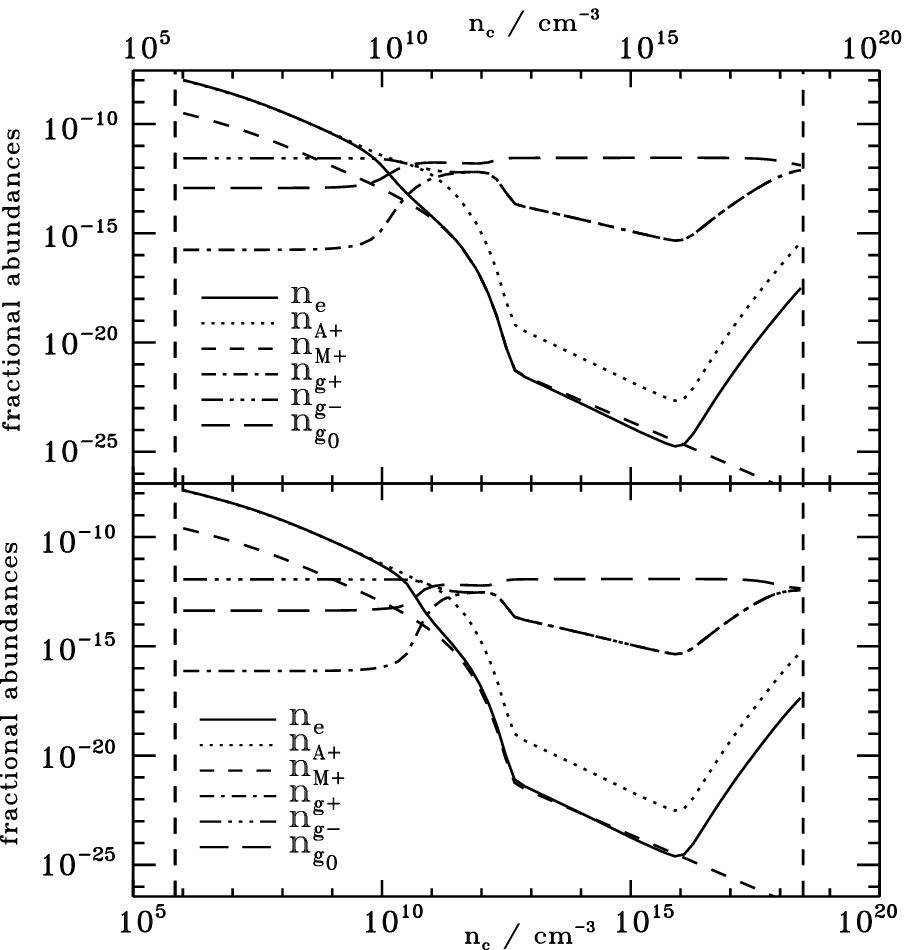}		
	\end{minipage}		
	\caption[Fractional abundances]
	         {Central fractional abundances of species. 
						\textbf{Top left:} Model 1, $a_{\mathrm{gr}}=0.019~\mathrm{\mu m}$,
						\textbf{Middle left:} Model 2, $a_{\mathrm{gr}}=0.038~\mathrm{\mu m}$,
						\textbf{Bottom left:} Model 3, $a_{\mathrm{gr}}=0.075~\mathrm{\mu m}$.				
					  The convergence of the abundances of positively- and negatively-charged grains
						is pushed to higher densities with increasing grain radius. The departure of the 
						electron abundance from the ion abundance also happens at higher densities with decreasing grain 
						surface area. The dashed vertical line at $n_{\mathrm{c}}\approx 10^{6}~\mathrm{cm}^{-3}$ indicates the density at which the chemistry 
						and non-ideal MHD treatment are switched on, while the one at $n_{\mathrm{c}}\approx 3\times 10^{18}~\mathrm{cm}^{-3}$ shows when 
						we assume the destruction of grains restores flux-freezing.
						\textbf{Top right:} Model 4, $a_{\mathrm{gr}}=0.113~\mathrm{\mu m}$,
						\textbf{Bottom right:} Model 5, $a_{\mathrm{gr}}=0.150~\mathrm{\mu m}$.
						Since the grain radius increases only by a factor of $4/3$ from Model 4, the effect on the abundances is less pronounced
						than it was for the smaller grains.}		
	\label{fig:abundances_DBK11}
\end{figure*}

In Figs. \ref{fig:ionfrac_DBK11} and \ref{fig:abundances_DBK11} we present the evolution of the central 
ionization fraction versus number density for the different models, and the abundances of grains and 
gas-phase species relative to molecular hydrogen. Figure \ref{fig:ionfrac_DBK11} shows that at low 
densities the smaller grains result in a lower ionization fraction. The reason is that small grains are 
highly abundant and thus have a large surface area to which the gas-phase charges can adhere. An increasing
density causes the ionization fraction to decrease with a slope slightly 
steeper than the canonical relation $\propto n_{\mathrm{n}}^{-1/2}$ for cosmic-ray-dominated ionization. 
At intermediate densities the decrease in ionization fractions level off, starting at the lowest density for the smallest 
grains and at progressively higher densities for successively larger grains. This phenomenon is driven 
by the electrons adsorbing to the grains. The ions also adsorb to the grains, but due to a lower 
thermal speed than the electrons, their collision rate is much smaller.   
With the grains acting as a reservoir for charges, and cosmic-ray 
ionization continuing to be active, the decrease in ionization fraction
is temporarily halted.
This occurs at later stages with increasing grain size, as 
Fig. \ref{fig:abundances_DBK11} shows, 
for the reason of the lower
overall grain abundance (scaling as $a_{\mathrm{gr}}^{-3}$), which cannot be compensated
by a larger collision rate (see Eq.(\ref{eq:alpha_ei_g}) in Appendix \ref{app:rate_coeffs}).
Beyond $n_{\mathrm{c}} \gtrsim 10^{12}~\mathrm{cm}^{-3}$, the ionization fraction in Fig. \ref{fig:ionfrac_DBK11} 
drops precipitously for all grain sizes, as cosmic rays become shielded due to high column densities.
The evolution at still higher densities is determined by radioactivity, until finally thermal ionization 
kicks in at $n_{\mathrm{c}} \approx 10^{16}~\mathrm{cm}^{-3}$. Potassium is one of the first 
species to be ionized, but as the temperature increases further, grain evaporation also 
releases charges into the gas. 

\begin{figure}[htp]%
    \includegraphics[width=0.9\hsize]{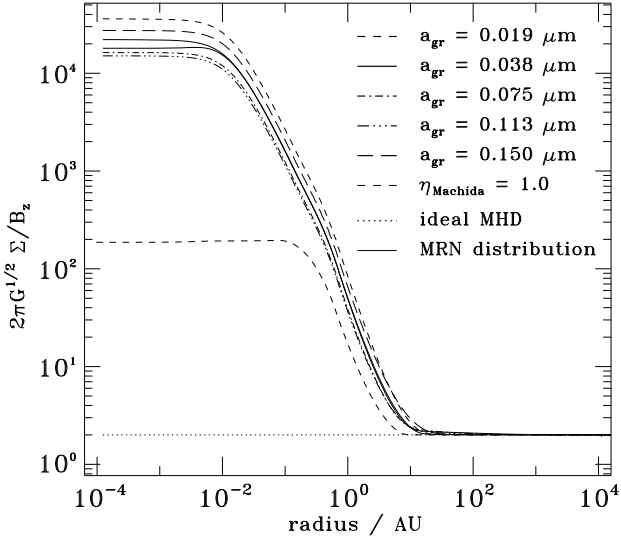}%
		\centering
    \caption[{Mass-to-flux ratio versus radius for different grain sizes.}]
             {Mass-to-flux ratio $\mu$ versus radius for different grain sizes at the time of the formation of the second core.
						  For comparison, the thin solid line shows the case with only the simplified version of Ohmic dissipation, 
							as used in DB10. In the center, the difference is a factor of $\approx 10^{2}$, but further out it
							it still $2\times$ higher, and also shows a greater extent by about this factor. The reason is twofold: firstly, 
							ambipolar diffusion is important at lower densities and, secondly, the effective resistivity is larger everywhere than the parametrization used in 
							DB10 (cf. Fig. \ref{fig:diffc_DBK11}).
						 }
    \label{fig:mu_DBK11}
\end{figure}

Figure \ref{fig:mu_DBK11} compares the mass-to-flux ratio (normalized to the value critical for collapse) 
for the various models at the time thermal ionization reestablishes flux freezing, when 
$n_{\mathrm{c}} \gtrsim 10^{19}~\mathrm{cm}^{-3}$, shortly before the formation of a central
protostar. For comparison, we also show the 
mass-to-flux ratio for a model with flux-freezing throughout, and the model with only the 
simple prescription for Ohmic dissipation used in DB10. The inclusion of 
ambipolar diffusion and the more realistic treatment of the microphysics 
lead to a further weakening of the magnetic field, both in the first core (by a factor of 
$\approx 10^{2}$), and in the region outside (by a factor of $\approx 2$). The region with 
increased mass-to-flux ratio also extends out slightly further, also by a factor of 
$\approx 2$. This is caused by 
the higher effective value of the resistivity and the earlier onset of its efficacy. 
The mass-to-flux ratio increases by a factor of $\approx 10^{4}$ over the course of
the evolution, to $\approx 20,000$ times the critical value. Field strengths
in main-sequence stars and T-Tauri stars require flux reduction of a factor $\approx 10^{4}-10^{8}$ 
(even if their field is dynamo-generated) if their parent molecular clouds are assembled 
from subcritical $\mu < 1$ atomic gas. This is known as the ``magnetic flux problem'' of 
star formation. Our results constitute substantial progress towards understanding its resolution.

\begin{figure}[htp]%
    \includegraphics[width=0.9\hsize]{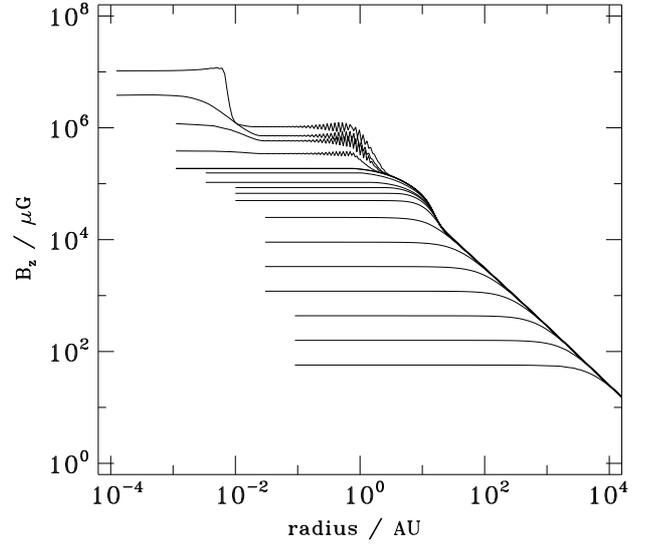}%
		\centering
    \caption[Evolution of $B_{z}$ versus radius over time]
            {Evolution of $B_{z}$ versus radius over time. The thin lines are plots at the times listed in Table \ref{table:times}, for 
						the fiducial grain radius of $a_{\mathrm{gr}}=0.038~\mathrm{\mu m}$.}
    \label{fig:b_z_time_DBK11}
\end{figure}

Figure \ref{fig:b_z_time_DBK11} shows the evolution of the radial profile of the vertical component of the magnetic field.
At low densities, 
$B_{z}$ increases under near flux freezing. Ambipolar diffusion is present and active, but is too slow to be dynamically 
important. Dramatic flux loss occurs once grains become the dominant charge carriers at 
$n \approx 10^{11}~\mathrm{cm}^{-3}$ and the field evolution slows. This can be seen in the bunching-up of the thin lines 
that are snapshots at different times (at constant increments of number density, at times given in Table \ref{table:times}). 
The small fluctuations near the interface of the first core result from differentiating $\eta _{\mathrm{eff}}$ and 
taking a second derivative of $B_{z}$ in the induction equation. 
Near the interface of the first core, the steep dependence of $\eta _{\mathrm{eff}}$ on density in the range
$10^{11}~\mathrm{cm}^{-3}<n<10^{13}~\mathrm{cm}^{-3}$ (see Fig. \ref{fig:diffc_agr_DBK11}) combined with an extreme gradient of the density, cause a jump
in the diffusion coefficient of many orders of magnitude. 

We stress that these fluctuations are not numerical instabilities and do not grow with time. The implicit, adaptive ODE solver we use to solve 
the MHD equations with the method of lines (see Section \ref{sec:method_DBK11}) is unconditionally stable.

\begin{figure}[htp]%
    \includegraphics[width=0.9\hsize]{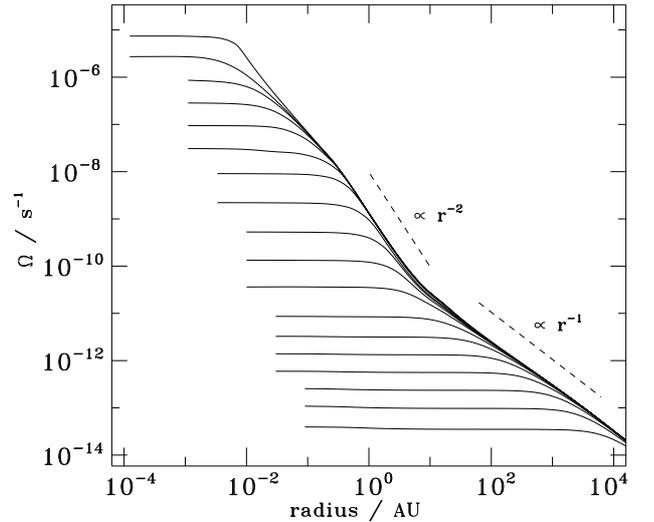}%
		\centering
    \caption[Evolution of $\Omega$ versus radius over time]
            {Evolution of the angular velocity $\Omega$ versus radius over time. The thin lines are plots at the times listed in Table \ref{table:times}, for 
		the fiducial grain radius of $a_{\mathrm{gr}}=0.038~\mathrm{\mu m}$. The profile changes from being $\propto r^{-1}$ to $\propto r^{-2}$ in the expansion 	
		wave, as expected from theoretical considerations.}
    \label{fig:omega_time_DBK11}
\end{figure}

In Fig. \ref{fig:omega_time_DBK11} we present the profile of the angular velocity at various times. 
At large radii, $\Omega \propto r^{-1}$ because the core evolves under near angular momentum conservation with the specific angular momentum $j \equiv \Omega r^{2} \propto m_{\rm encl}$ and $\Sigma
\propto r^{-1}$ in the prestellar infall (cf. Fig. \ref{fig:sigma_time_DBK11}).
Inside the expansion wave, there is a break in the angular velocity profile, as expected 
from theoretical considerations \citep[e.g.,][]{SaigoHanawa1998}. The enclosed mass in that region is essentially the mass of
the first core (the expansion wave itself does not contribute much), and so is approximately constant with radius. 
Then $j \propto m_{\mathrm{encl}} \approx
\mathrm{const}$, and therefore $\Omega \propto r^{-2}$.

\begin{figure}[htp]%
    \includegraphics[width=0.9\hsize]{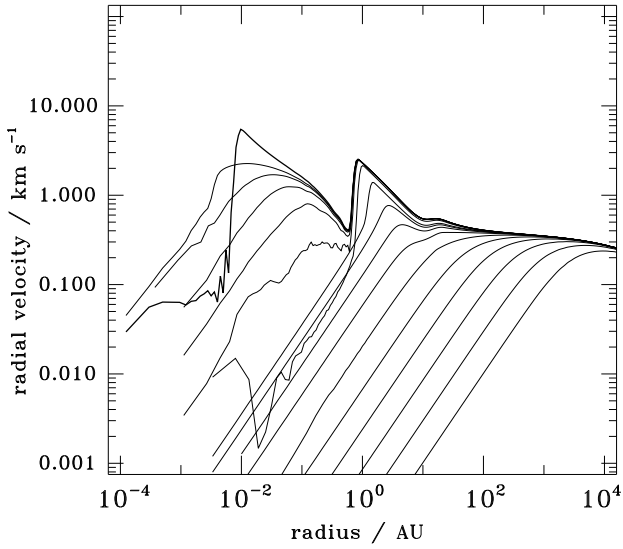}%
		\centering
    \caption[Evolution of $v_{r}$ versus radius over time]
            {Evolution of the infall velocity velocity $\left\vert v_{r} \right \vert$ versus radius over time. 
	     The thin lines are plots at the times listed in Table \ref{table:times}, for the fiducial grain radius of 
	     $a_{\mathrm{gr}}=0.038~\mathrm{\mu m}$. Outside the first core, the magnetic wall causes a small bump. 
	     Within the first and second cores material is moving about inwards 
	     and outwards with small speeds. This causes spikes when absolute values are plotted on double-logarithmic axes.
	     }
    \label{fig:vr_time_DBK11}
\end{figure}

The radial velocity (Fig. \ref{fig:vr_time_DBK11}) shows the first and second cores very clearly. At their edges 
($\approx 1~\mathrm{AU}$ and $\approx 10^{-2}~\mathrm{AU}=2~\Rsun$, respectively), accretion
shocks develop and the velocity drops precipitously. Outside the cores, in the expansion wave, 
the velocity follows a power law $\propto r^{-1/2}$. At $\approx 10~\mathrm{AU}$, 
a slight bump in the infall velocity hints at the magnetic wall. The fluctuations within the first and
second cores stem from the fact that we plot absolute values in a log-log plot: in nearly-stable 
conditions as prevalent there, the velocity can be positive or negative, but remains small.

\subsection{Disk formation}\label{subsec:results_disk}

The second core is a truly hydrostatic object, and very nearly spherical, and so the 
thin-disk approximation breaks down there. Therefore, once the central number density 
$n_{\mathrm{c}} \approx 10^{20}~\mathrm{cm}^{-3}$, we introduce a sink cell of radius 
$\approx 3~\Rsun$. Note that this is two orders of magnitude smaller than is commonly 
used \citep[e.g.,][]{MellonLi2008,MellonLi2009}.

\begin{figure}[htp]%
    \includegraphics[width=0.9\hsize]{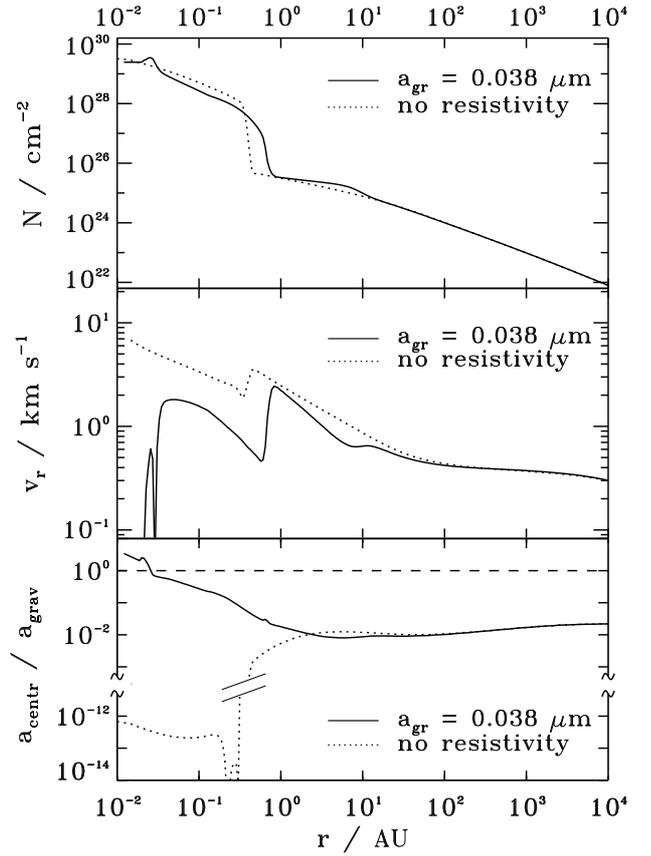}%
		\centering
    \caption[Disk formation]
            {Evidence for disk formation. \textbf{Top panel:} column density profile. The system is Toomre-unstable, 
		and breaks up into a ring in the resistive model (solid line), but not in the flux-frozen model
		(dotted line). \textbf{Middle panel:} infall profile. Infall is stopped in the resistive model (solid line), in innermost region where
a centrifugal disk forms, but not in the flux-frozen model (dotted line) \textbf{Bottom panel:} ratio
		of centrifugal over gravitational acceleration. In the resistive model, centrifugal balance is achieved. 
		In contrast, catastrophic magnetic braking occurs for the flux-frozen model, the centrifugal support drops 
		to negligible values and the first core is spun down drastically.
		}
    \label{fig:disk_plot}
\end{figure}

In Fig. \ref{fig:disk_plot} we present evidence for the formation of a centrifugal disk. The figure shows the profiles of column density, infall velocity, and the ratio of centrifugal to gravitational acceleration shortly after the introduction of the sink cell. This is done for Model 2, which has resistivity and the fiducial grain size $a_{\mathrm{gr}}=0.038~\mathrm{\mu m}$, and for a model without resistivity. In the resistive model, centrifugal balance is achieved in a small region ($\approx 10~\Rsun$) close to the center (bottom panel), while the flux-freezing model is braked ``catastrophically'', and the support drops to minuscule values. Centrifugal balance is a necessary and sufficient condition for the formation of a bona-fide centrifugally-supported disk (as opposed to a larger flattened structure that is still infalling). At the same time all infall is halted there and the radial velocity plummets (middle panel), while in the flux-freezing model infall continues unhindered. After a few more months of evolution, a Toomre instability develops, and the rotationally-supported structure breaks up into a ring (top panel, solid line). At this point, we stop the simulation, because more physics would be required to follow the further evolution of the disk. 

DB10 also presented a model without magnetic braking (their Fig. 1). In that case, the entire first core turns into a large centrifugally-supported structure of size several $\mathrm{AU}$. 

\begin{figure*}
  \centering
	\begin{minipage}[b]{0.46\hsize}		
		\centering
		\includegraphics[scale=0.98]{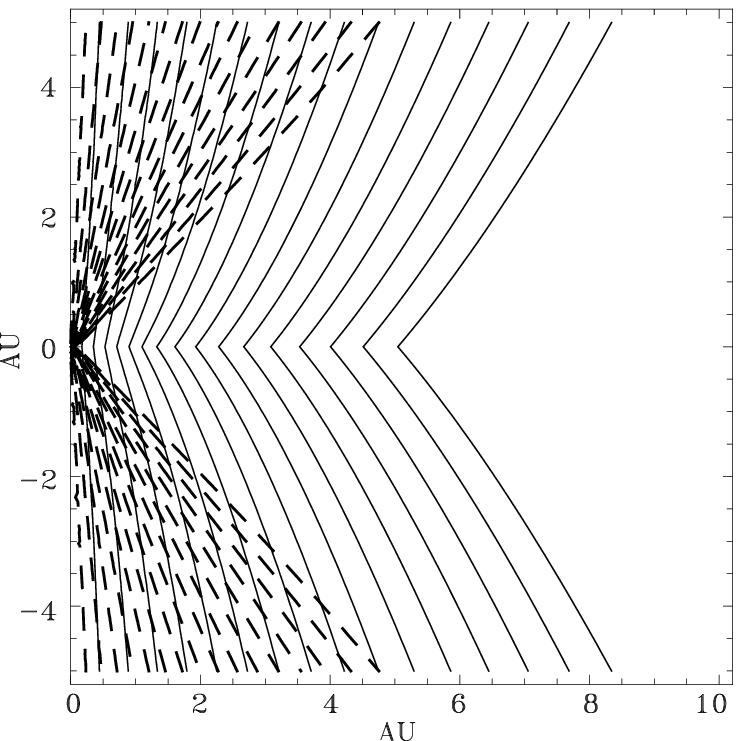}		
	\end{minipage}
	\hspace{0.5cm}
	\begin{minipage}[b]{0.46\hsize}	
		\centering
		\includegraphics[scale=0.98]{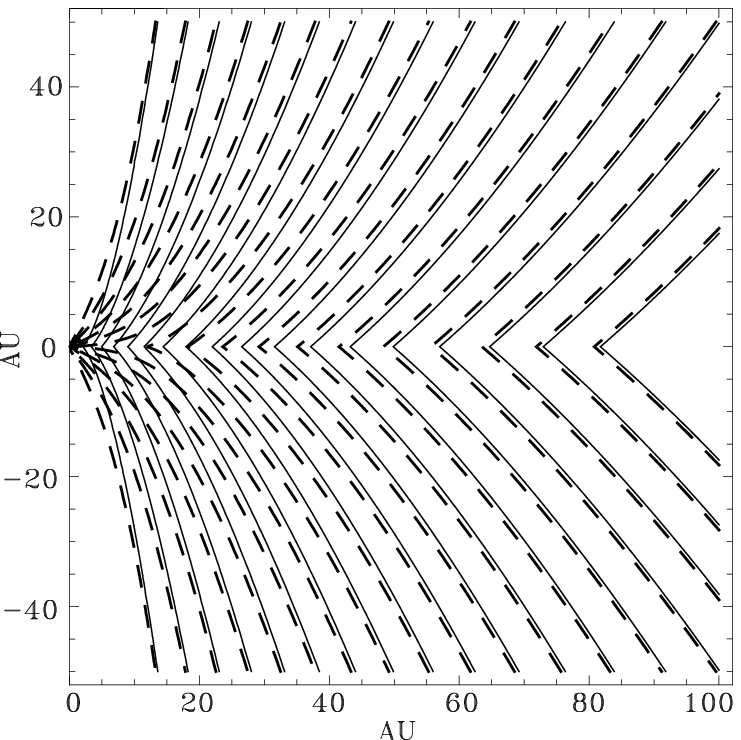}		
	\end{minipage}		
	\caption[Magnetic field lines on $10$ and $100~\mathrm{AU}$ scales]
	        {Magnetic field lines. The box on the left has dimensions $10~\mathrm{AU}$ on each side, 
					 while the box on the right has dimensions $100~\mathrm{AU}$. The dashed lines represent 
					 the flux-freezing model, while the solid lines show \textit{the same} field lines for 
					 the model including non-ideal MHD effects for a grain size 
					 $a_{\mathrm{gr}}=0.038~\mathrm{\mu m}$. In both cases, the second core has just formed 
					 and is on the left axis midplane. The field lines straighten out significantly on small scales
					 in the non-ideal model compared to the flux-frozen model.
					}		
	\label{fig:fieldlines_DBK11}
\end{figure*}

Figure \ref{fig:fieldlines_DBK11} shows the magnetic field line topology above and below the disk on two scales ($10~\mathrm{AU}$ and $100~\mathrm{AU}$) for both the flux-freezing and non-ideal MHD models. The field lines are calculated shortly after the formation of the second core, assuming force-free and current-free conditions above a finite thin disk \citep{MestelRay1985}. The split monopole of the flux-frozen model (dashed lines) is created as field lines are dragged in by the freely falling material within the expansion wave front at $\approx 15~\mathrm{AU}$. This is replaced by a more relaxed field line structure in the non-ideal case (solid lines), for which the field is almost straight in the central region. \citet{GalliEtAl2009} presented similar field configurations resulting from a simplified model for Ohmic dissipation during the collapse. The extreme flaring of field lines in the flux-freezing model is a fundamental cause of the magnetic braking catastrophe. 

\begin{figure}[htp]%
    \includegraphics[width=0.9\hsize]{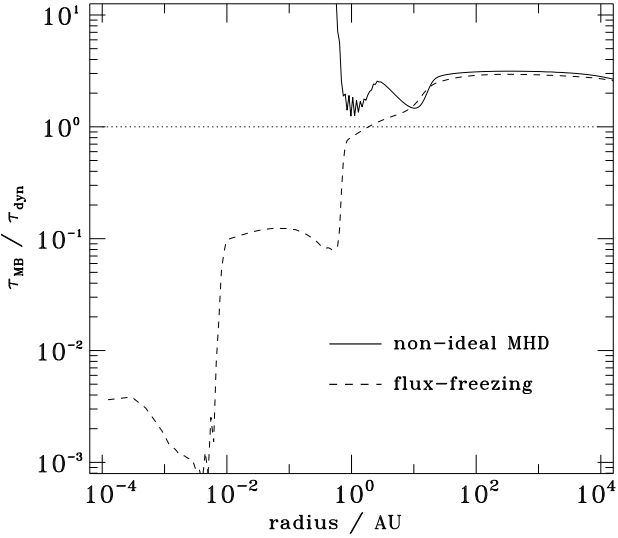}%
		\centering
    \caption[Magnetic braking condition]
            {Plot of the ratio $\tau _{\mathrm{MB}} / \tau _{\mathrm{dyn}}$. 
		For both the flux-freezing case (dashed line) as well for non-ideal MHD (solid line), the magnetic braking
		time is reduced in the expansion wave region from what is was in the prestellar infall region. In the case of non-ideal
		MHD, it turns up sharply at the interface of the first core where diffusion has reduced the magnetic field strength
		significantly and rendered magnetic braking ineffective. In the flux frozen case however, the first core
		marks a sharp drop in magnetic braking time to only a tenth of the dynamical time. 
						}
    \label{fig:MB_cond_DBK11}
\end{figure}

We can estimate the efficiency of magnetic braking by comparing its instantaneous timescale with that of the dynamical evolution of the core. The
relevant ratio is 
\begin{equation}
    \frac{\tau _{\mathrm{MB}}}{\tau _{\mathrm{dyn}}} = \frac{L / N_{\mathrm{cl}}}{r/v_{r}}, 
\label{eq:MB_condition}
\end{equation}
where $L = \Sigma \Omega r^{2}$ is the angular momentum per unit area, $N_{\mathrm{cl}} = r B_{\varphi} B_{z} / 2\pi $ is the torque per unit area acting on the cloud, and $\tau _{\mathrm{dyn}} = r/v_{r}$ is the dynamical time. Figure \ref{fig:MB_cond_DBK11} shows this ratio for catastrophic magnetic braking for our cloud after the second core has formed. In the region of dynamical collapse still characterized by a prestellar infall profile, magnetic braking is not very effective (its time scale being $\approx 3$ times that of dynamic evolution) and the cloud evolves under approximate angular momentum conservation. In the non-ideal MHD case (solid line), the magnetic braking time is never shorter than the dynamical time, and thus catastrophic magnetic braking is avoided. However, the ratio comes close to unity in the expansion wave where a significant amount of angular momentum loss can still occur. Within the first core however, where the magnetic field has been weakened by diffusive effects, magnetic braking is totally disabled, and the ratio $\tau _{\mathrm{MB}}/\tau _{\mathrm{dyn}}$ rises dramatically.  For flux freezing (dashed line), the ratio drops below unity within the expansion wave, and the first core marks a further sharp drop in the magnetic braking time to only a tenth of the dynamical time. Both the decline of $\tau _{\mathrm{MB}}/\tau _{\mathrm{dyn}}$ within the expansion wave as well as its sharp drop within the first core can be considered elements of catastrophic
magnetic braking.

\ctable[
cap         =  Scaling laws in the flux-freezing case.,
caption     =  Scaling laws in the flux-freezing case.,
label       =  tab:scaling_laws,
mincapwidth =  0.8\hsize,
]{lll}{}{ \FL
                     &  prestellar profile & expansion wave     \ML
$\Sigma$             & $\propto r^{-1}$    & $\propto r^{-1/2}$ \NN 
$\varrho$            & $\propto r^{-2}$    & $\propto r^{-1}$   \NN
$v_{r}$              & $\approx$ const     & $\propto r^{-1/2}$ \NN
$g_{r}$              & $\propto r^{-1}$    & $\propto r^{-2}$   \NN
$\Omega$             & $\propto r^{-1}$    & $\propto r^{-2}$   \NN
$\Phi                $ & $\propto r$       & $\approx$ const    \NN
$m_{\mathrm{encl}}$  & $\propto r$         & $\approx$ const    \NN
$B_{z}$              & $\propto r^{-1}$    & $\propto r^{-1/2}$ \NN
$B_{r}$              & $\propto r^{-1}$    & $\propto r^{-2}$   \NN
$B_{\varphi}$        & $\propto r^{-1}$    & $\propto r^{-2}$   \LL
}

In the flux-freezing case (where the magnetic field eventually forms essentially a split-monopole configuration, dashed line of Fig. \ref{fig:fieldlines_DBK11}), simple analytic scaling arguments can be made to explain the profiles. These are summarized in Table \ref{tab:scaling_laws}. The vertical component of the magnetic field, $B_{z}$, always scales like the column density $\Sigma$ due to the spatially constant mass-to-flux ratio, and magnetic flux conservation (flux-freezing), while $B_{r}$ scales like the gravitational field. The azimuthal component of the magnetic field, $B_{\varphi} \propto \Phi \Omega r^{-1}$ from the magnetic braking calculation (see Section \ref{sec:magn_braking}). 

In the prestellar infall region under flux freezing, the magnetic braking time 
\begin{align}
  \tau _{\mathrm{MB}} &= \frac{L}{N_{\mathrm{cl}}} \propto \frac{\Sigma \Omega r^{2}}{B_{z} B_{\varphi} r} 
\propto \frac{\Sigma r^{2}}{B_{z} \Phi}
	                     \propto r.
\label{eq:tau_MB_scaling_prestellar}
\end{align}
Similarly, the dynamical time in the prestellar profile scales as
\begin{align}
  \tau _{\mathrm{dyn}} &= \frac{r}{v_{r}} \propto r
\label{eq:tau_dyn_scaling_prestellar}
\end{align}
since $v_r \approx$ constant.
The magnetic braking time and the dynamical time therefore scale with radius in the same way; in other words their ratio is constant. 

In the expansion wave region, the enclosed flux is dominated by the split-monopole, and is approximately constant (similar arguments hold for the enclosed mass). Furthermore, $\Sigma$, $B_{z}$, and $v_r$ scale differently than in the region of prestellar infall (see Table \ref{tab:scaling_laws}). Then in the expansion wave region
\begin{align}
  \tau _{\mathrm{MB}} &\propto \frac{\Sigma r^{2}}{B_{z} \Phi} \propto r^{2}, \\
	\tau _{\mathrm{dyn}} &= \frac{r}{v_r}  \propto r^{3/2}.
\label{eq:tau_scaling_expansion}
\end{align}
Here, the ratio of magnetic braking time and dynamical time is $\propto r^{1/2}$, as seen in Fig. \ref{fig:MB_cond_DBK11} between $\approx 1~\mathrm{AU}$ and $\approx 15~\mathrm{AU}$ (for the dashed line).
This shows that the magnetic braking catastrophe is a fundamental 
property of the expansion wave in a flux-frozen model.
In the non-ideal case, the magnetic field does not assume a split-monopole configuration, and neither does the mass-to-flux ratio remain spatially constant, and simple scaling laws are not applicable.

\begin{figure}[htp]%
    \includegraphics[width=0.9\hsize]{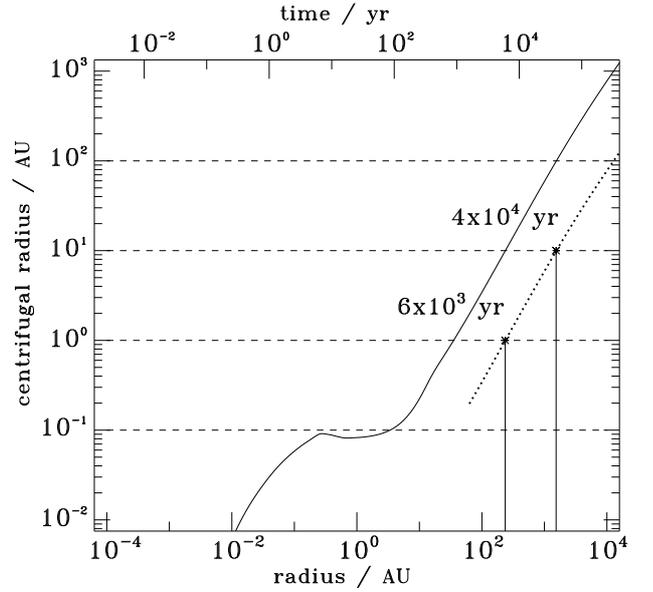}%
		\centering
    \caption[Specific centrifugal radius]
            {Estimated centrifugal radius of each mass shell. Beyond $\approx 10~\mathrm{AU}$, the
						centrifugal radius increases linearly. The solid line is the current centrifugal radius, while the
						dotted line is an estimate of $r_{\mathrm{cf}}$ after the expansion wave has passed and the 
						associated magnetic braking has reduced the specific angular momentum (see text). The top axis shows
						the approximate times at which the expansion wave is expected to pass a certain radius. The symbols 
						show the estimated centrifugal radii of $1~\mathrm{AU}$ and $10~\mathrm{AU}$.
						The associated times are $6\times 10^{3}~\mathrm{yr}$ and $4\times 10^{4}~\mathrm{yr}$.

						}
    \label{fig:r_cf_DBK11}
\end{figure}

The centrifugal radius of a mass shell, i.e., the radius at which the material would achieve centrifugal balance with gravity if angular momentum were
conserved, can be estimated according to 
\begin{equation}
    r_{\mathrm{cf}}=j^2/GM, 
\label{eq:r_cf}
\end{equation}
where $j=\Omega r^{2}$ is the specific angular momentum, and $M$ is the mass in the 
central object. While this expression is strictly true for infall onto a point mass, 
we take $M$ to be the enclosed mass. The result is shown in Fig. \ref{fig:r_cf_DBK11}. 
This estimation is imperfect within the first core, but well-justified outside it 
since the gravitational potential there looks like that of a point mass. At larger 
radii, most of the enclosed mass is in the cloud and not the central mass, but by the 
time the expansion wave reaches those mass shells and rapid infall begins, approximately 
half of the enclosed mass will be in the compact central object or disk
\citep[e.g.,][]{Shu1977}. The solid line is the estimate based on Eq.(\ref{eq:r_cf}) but 
the dotted line takes into account an estimated loss of specific angular momentum $j$ by 
a factor of $3-4$ in the expansion wave, so
that $j^2$ is reduced by a factor of 10. This estimate is supported by
data presented later in Fig. \ref{fig:spec_ang_mom_encl_mass} and is applied to mass shells that are well outside the expansion wave at the time that we stop our simulation. 
The dotted line may represent
an {\it upper limit} to the disk radius since some more angular momentum
may be lost by magnetic braking when the disk is forming, if magnetic
field dissipation has not completely shut off magnetic braking by that 
time. Future calculations can settle that issue. Outflows will also remove
some angular momentum from the forming disk, reducing the size further, 
although a counter-effect that we do not quantify is disk expansion due to 
internal angular momentum transfer within the disk. We assume the latter to not be 
significant in the early Class 0 phase.  
The dotted line reveals a
centrifugal radius $r_{\mathrm{cf}} \approx 1~\mathrm{AU}$ for mass shells currently at a distance of $200~\mathrm{AU}$ from the center, and $r_{\mathrm{cf}} \approx 10~\mathrm{AU}$ for mass shells at $1,500~\mathrm{AU}$.  
The upper axis labels the times at which the expansion wave (traveling 
at the sound speed) is estimated
to reach these radii, and rapid infall to begin. The subsequent infall
to the disk should take significantly less time than the expansion
wave arrival time, since the infall is highly supersonic.
Hence, the expansion wave arrival times can be taken to be approximate
estimates of the central object age just as the centrifugal radii
estimates can be taken to be upper limits to the disk size.

\begin{figure}[htp]%
    \includegraphics[width=0.9\hsize]{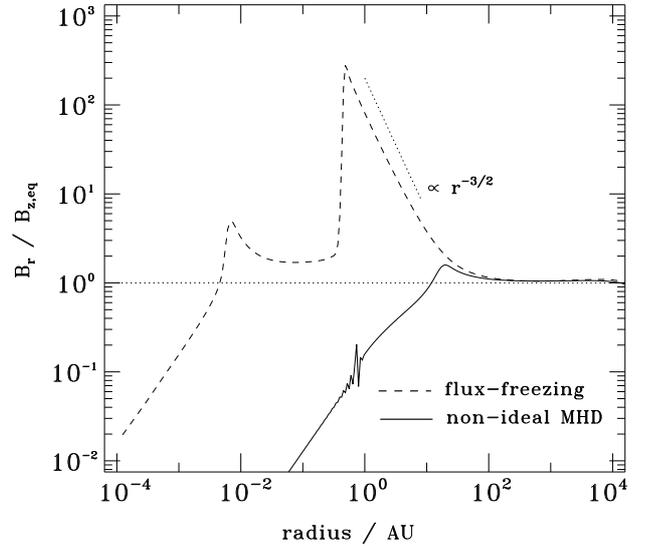}%
		\centering
    \caption[Ratio of magnetic field components]
            {Ratio of $B_{r}/B_{z,\mathrm{eq}}$. In the flux frozen case (dashed line), $B_{r}$ dominates everywhere within the expansion wave, 
		indicative of extreme flaring and a split-monopole field configuration. With field diffusion active (solid line), the field lines straighten out quickly 
		inside the expansion wave (cf. Fig. \ref{fig:fieldlines_DBK11})
		}
    \label{fig:b_z_b_r_comparison}
\end{figure}

\subsection{Further results}

Figure \ref{fig:b_z_b_r_comparison} shows the ratio of the radial and vertical components of the magnetic field, $B_{r}/B_{z,\mathrm{eq}}$. 
This is a quantitative measure of the local bending of the magnetic field (while Fig. \ref{fig:fieldlines_DBK11} is a visual representation). 
For flux freezing (dashed line), $B_{r}$ dominates everywhere within the expansion wave, which is indicative of the extremely-flared 
split-monopole field configuration. Foot points of magnetic flux tubes are connected to head points at much larger radii. This large lever 
arm leads to rapid angular momentum loss --- the magnetic braking catastrophe \citep[][]{GalliEtAl2009}. In the diffusive case (solid line), 
the field lines straighten out quickly inside the expansion wave, and only a small increase in pitch angle occurs inside it. 

Again, we can make scaling arguments in the flux freezing case. The radial component of the 
magnetic field, $B_{r}$, always scales like the radial gravitational field. In the prestellar 
region of the profile (outside $\approx 100~\mathrm{AU}$) $B_{r}$ and $B_{z}$ both scale 
$\propto r^{-1}$ and so their ratio is constant. In the expansion wave outside the first core 
(between $\approx 1~\mathrm{AU}$ and $\approx 15~\mathrm{AU}$) on the other hand, 
$B_{r} \propto r^{-2}$ while $B_{z} \propto \Sigma \propto r^{-1/2}$. Their ratio is therefore 
$\propto r^{-3/2}$. As before, the magnetic field does not assume a split-monopole configuration 
in the non-ideal case, and simple scaling laws are therefore not applicable.

\begin{figure}[htp]%
    \includegraphics[width=0.9\hsize]{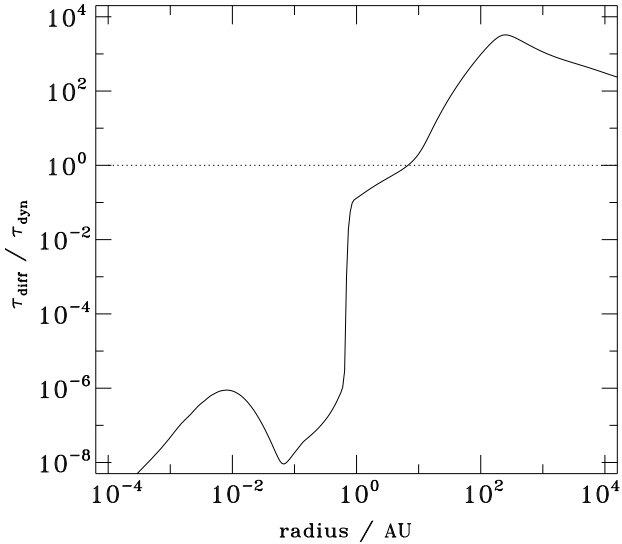}%
		\centering
    \caption[Diffusion condition]
            {Ratio of diffusion time $\tau _{\mathrm{diff}}$ and dynamical time $\tau _{\mathrm{dyn}}$ at the end of the run. Within the expansion wave, the ratio is smaller than unity, 
						and diffusion dominates over advection. 
						}
    \label{fig:tau_diff_tau_dyn}
\end{figure}

Figure \ref{fig:tau_diff_tau_dyn} shows the ratio of diffusion time $\tau _{\mathrm{diff}}$ and dynamical time $\tau _{\mathrm{dyn}}$ 
near the end of our simulation run. \citet{ContopoulosEtAl1998} showed from scaling arguments that upon close approach to the protostar, 
diffusion of magnetic flux will eventually dominate advection. For our model, this is already the case in the expansion wave outside the 
first core. The reason for the sudden drop in the ratio at the interface of the first core is that the diffusion time scale 
$\tau _{\mathrm{diff}} \propto L^{2}/\eta _{\mathrm{eff}}$, where $L$ is the local scale length. However, not only does the density at 
this location increase very steeply, $\eta _{\mathrm{eff}}$ itself increases very steeply in the relevant density range too (see Fig. 
\ref{fig:diffc_DBK11}). Though mitigated by a simultaneous decrease in the dynamical time scale (see Fig. \ref{fig:vr_time_DBK11}), 
it still effects a drop of 5 orders of magnitude in the time scale ratio.

\begin{figure}[htp]%
    \includegraphics[width=0.9\hsize]{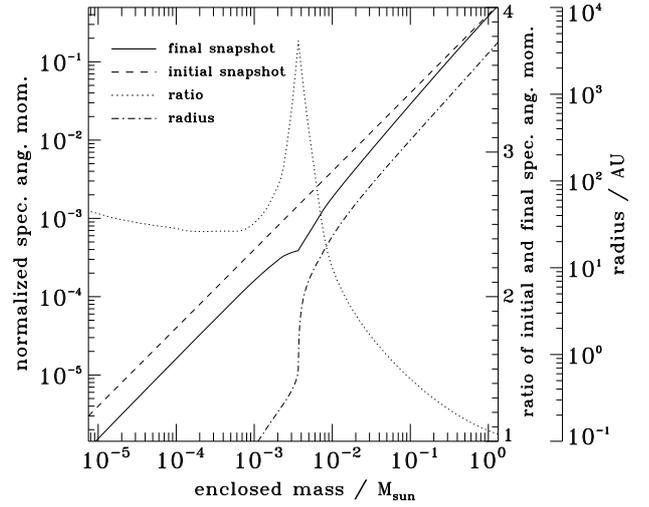}%
		\centering
    \caption[Specific angular momentum]
            {Specific angular momentum versus enclosed mass at initial state (dashed line) and final snapshot (solid line), and their ratio (dotted line). 
						The expansion wave brings with it a maximal reduction of specific angular momentum. Further insight is provided by the radius versus enclosed mass, 
						plotted as the dot-dashed line, with radius values denoted on the second axis on the right.						
						}
    \label{fig:spec_ang_mom_encl_mass}
\end{figure}

In Fig. \ref{fig:spec_ang_mom_encl_mass}, we plot the specific angular momentum versus enclosed mass at the initial state (dashed line) 
and the final snapshot (solid line), as well as their ratio (dotted line). The expansion wave brings with it a maximal reduction 
(a factor of $\lesssim 4$) of specific angular momentum. The radius versus enclosed mass is plotted as the dot-dashed line.	The steep 
rise of the latter at $\approx 4 \times 10^{-3}~\Msun$ is a consequence of the first core dominating the enclosed mass, and the material 
within the expansion wave only slowly increasing it at larger radii.

\section{Discussion}\label{sec:Discussion}

Our results lead us to propose the following scenario of disk formation and evolution in low-mass stars: 
\begin{enumerate}
	\item A small disk forms, initially $\ll 1~\mathrm{AU}$, but eventually encompasses the entire first 
core with $\approx 1~\mathrm{AU}$, since magnetic diffusion is very efficient there and inactivates magnetic braking. 
	\item Figure \ref{fig:r_cf_DBK11} shows that the estimated centrifugal radius of a mass shell at radius $\lesssim 50~\mathrm{AU}$ lies 
	      within $\approx 1~\mathrm{AU}$. That means that the matter can fall to this radius without hitting a rotational barrier.
	\item At the same time, Fig. \ref{fig:MB_cond_DBK11} shows that magnetic braking is active within the expansion wave, while it is dormant 
	in the region of dynamical infall further out. Any material within the expansion wave (which moves outward at the local speed of sound) 
	will lose part of its angular momentum by magnetic braking (typically a factor of $3-4$, cf. Fig. \ref{fig:spec_ang_mom_encl_mass}), and 
	have its centrifugal barrier moved further inward. This is reflected in the dotted line in Fig. \ref{fig:r_cf_DBK11}, which estimates the 
	final centrifugal radius of a mass shell. After the expansion wave passes, material as far as $r \approx 200~\mathrm{AU}$ can directly 
	accumulate onto the inner $\approx 1~\mathrm{AU}$. Furthermore, a centrifugal radius $\lesssim 10~\mathrm{AU}$ is expected for gas currently 
	within $\approx 1500~\mathrm{AU}$, which encloses about $0.4~\Msun$ in our model. The expansion wave will take 
	$\approx 4\times 10^{4}~\mathrm{yr}$ to reach that point. We therefore predict no disk larger than $\approx 10~\mathrm{AU}$ to be detectable 
	in Class 0 objects younger than $\approx 4\times 10^{4}~\mathrm{yr}$. This agrees well with the observations of \citet{MauryEtAl2010} who do 
	not find evidence for disks $>50~\mathrm{AU}$ in Class 0 objects. ALMA will soon allow observers to test this prediction more stringently.
	\item The material in the disk will be subjected to internal mechanisms of angular momentum redistribution, e.g., the MRI 
	\citep[][]{BalbusHawley1991,Balbus2009}, and/or gravitational torques. It is a fair assumption that a disk in which self-gravity is important 
	self-regulates to $Q \approx 1$ by such processes \citep[e.g.,][]{VorobyovBasu2007}. The parameter 
	$Q= c_{\mathrm{s}}\Omega/\left(\pi G \Sigma\right)$ approximately determines whether a rotating disk is unstable to fragmentation 
	\citep{Toomre1964,GoldreichLyndenBell1965}. The exact critical value depends on the geometry and the equation of state, but generally spirals 
	and clumps will form if $Q\lesssim 1$, while a situation with $Q \gtrsim 1$ is stable \citep[e.g.,][]{Gammie2001}.
	\item Redistribution of angular momentum allows material in the inner disk to be funneled onto the star, while material further out gains the 
	excess angular momentum and expands its orbit. \citet{Basu1998} showed that internal redistribution of angular momentum means that the disk radius obeys
	\begin{equation}
	    r_{\mathrm{final}}\simeq 	r_{\mathrm{initial}}\left( \frac{M_{\bigstar}}{M_{\mathrm{disk}}}\right)^{2},
\label{eq:r_final}
	\end{equation}
	where $M_{\mathrm{disk}}$ and $M_{\bigstar}$ are the disk and star mass, respectively. This means that the disk will expand by a factor of $10^{2}$ by the time that $90\%$ of its mass has accreted onto the central object.	
	\item By the time the central object has accreted a significant amount of the available material (several $10^{5}~\mathrm{yr}$), a tenuous disk of several hundred AU will be present, consistent with observations \citep[e.g.,][]{AndrewsWilliams2005}. 
We expect that the disk will expand due to internal angular momentum 
redistribution, but that it will occur primarily after the Class 0 phase. During the Class 0 phase, the disk mass may remain a fair fraction of the star mass \citep{Vorobyov2011}, limiting the expansion implied by Eq.(\ref{eq:r_final}).
\end{enumerate}

The first step of the above scenario is broadly consistent with the models of \citet{MachidaMatsumoto2011} that have a small initial 
rotation rate. They have studied disk formation with their three-dimensional models that resolved the second core using a simplified 
resistivity. However, their models with rotation comparable to observations achieve a very rapidly rotating first core, which evolves 
directly into a 
``Keplerian'' (their words) disk before the central stellar core has been established. 
While the term ``Keplerian'' is normally reserved for when 
the central point mass 
dominates the overall gravitational potential, we consider that in their
context it refers to centrifugal balance.
In fact, we did obtain a similar result in DB10 for a model
with no magnetic braking, where the first core formed with rapid rotation that prevented
it from collapsing further. In the model of
\citet{MachidaMatsumoto2011} a stellar core does eventually form, but is surrounded by a Keplerian disk from the start. 
Conversely, in our models with an observationally-motivated rotation rate, the sustained effect of magnetic braking 
throughout the collapse ensures (even in the presence of Ohmic dissipation and ambipolar diffusion) that the first core 
can undergo rapid infall onto the second core. A centrifugally-supported disk is only built up subsequently through accretion.
We start with a differentially-rotating initial state, consistent with 
prior gravitational collapse \citep{Basu1997}, while 
\citet{MachidaMatsumoto2011} start with solid body rotation, which puts 
more angular momentum in the outer regions.
The two possible outcomes of direct second collapse followed by disk formation by accretion versus a large centrifugally-supported disk-like first core that coexists with a second core may certainly be a function of the implementation of magnetic braking and non-ideal MHD, initial rotation profiles, and the varieties of dimensionality and boundary conditions of different models. Future modeling can clarify this interesting duality of outcomes 
in the earliest phase of disk formation. 

Previous ideal and non-ideal MHD simulations of protostellar collapse 
focused on the formation of large centrifugal disks of size $\approx 100~\mathrm{AU}$. For example, \citet{MellonLi2009} and \citet{LiKrasnopolskyShang2011} find that the inclusion of ambipolar diffusion, Ohmic dissipation, and even the Hall term are not sufficient to avoid catastrophic magnetic braking at the distances $> 7$ AU from the origin that they model, for models with realistic initial magnetic field strength. \citet{MachidaEtAl2011} claim that a disk can finally form in later stages when the envelope is depleted and magnetic coupling to outer moment of inertia is therefore weak. \citet{KrasnopolskyEtAl2010} invoked ``anomalous'' resistivity to weaken the effects of magnetic braking and thereby allow for the formation of large centrifugal disks.

We note that there is no \textit{a priori} reason why centrifugal 
disks should be large in the early accretion phase. Using an observationally 
motivated magnetic-field strength and without recourse to ``anomalous'' 
resistivity, we have demonstrated that centrifugal disks \textit{can} form 
during the Class 0 stage, albeit not as large as $\approx 100~\mathrm{AU}$. 
Our result is made possible by our resolution of the innermost regions of collapse, and 
a sufficiently accurate treatment of the relevant chemical and magnetohydrodynamical processes. 
While not yet computationally feasible, well-resolved three-dimensional 
simulations that take into consideration these processes would represent 
a significant advance beyond the thin-disk approximation we have employed here.
 
The thin-disk approximation does not allow us to follow jets or outflows. While those effects undoubtedly play 
some role in removing angular momentum from protostellar cores \citep[e.g.,][]{PudritzNorman1983}, 
they do not start to affect the angular momentum balance until a centrifugal disk is present. Angular momentum 
loss due to outflows \textit{after} disk formation 
only strengthens our expectation that observations will not reveal any disks larger than $10~\mathrm{AU}$ in the 
early Class 0 phase. The disk would be even smaller than predicted by Eqns.(\ref{eq:r_cf}) and (\ref{eq:r_final}) if additional angular momentum and mass were extracted by outflows.

\section{Summary and conclusions}\label{sec:Summary_DBK11}

We present results from a new axisymmetric code using the thin-disk approximation that calculates the collapse of rotating magnetized prestellar cores. 
While treating vertical ($z$) structure and angular momentum transport in an approximate manner, our code allows us to 
follow the evolution all the way to stellar sizes and near-stellar densities. We determine the abundances of seven different species, 
and consider inelastic collisions. Our effective resistivity includes the effect of ambipolar diffusion and Ohmic 
dissipation for five different (single) grain sizes as well as for an MRN distribution of grain sizes. 

By following the complex relationships between the many nonlinear processes at work in the formation of stars (e.g. self-gravity, rotation, 
chemistry, thermodynamics, non-ideal MHD, grain effects, etc.), we have been able to simultaneously address both the angular momentum problem 
and the magnetic flux problem of star formation. These two processes are intricately and fundamentally linked.

We demonstrate the formation of a small centrifugally supported disk despite the effects of magnetic braking. The ``magnetic braking catastrophe'' 
is averted on small scales by substantial magnetic flux loss from the high-density region of the first core. This weakens the magnetic field, preventing it from 
spinning down the material in that region. The central mass-to-flux ratio increases by a factor of $\sim 10^{4}$. Shortly after the second collapse, 
disk formation happens very close to the central object, before the protostar has accreted a lot of mass ($M_{\bigstar} <  10^{-2}~\Msun$). This is consistent with the 
observational evidence of outflows at a very young age and the simultaneous non-detection of disks $\gtrsim 50~\mathrm{AU}$ around Class 0 objects \citep{MauryEtAl2010}. 
ALMA will allow observers to improve upon these constraints, by probing for disks of size $\approx 10~\mathrm{AU}$. 

We propose a disk formation scenario in which centrifugal disks form at the earliest stage of star formation and on small scales. 
We calculate the centrifugal radius 
for mass shells that are still infalling at the end of our simulation,
and estimate the disk to  
remain $\lesssim 10~\mathrm{AU}$ for $\approx 4 \times  10^{4}~\mathrm{yr}$.
In this case, a disk would not be observable around a young Class 0 object, even with ALMA. However, indirect indications of the 
presence of disks --- such as outflows --- can still be expected. The accretion disk slowly grows by continued accretion but more 
so by internal angular momentum redistribution, and may eventually reach the size $\approx 100~\mathrm{AU}$ observed around Class II objects.

\section*{Acknowledgments}
The authors thank Jan Cami for providing computational facilities to run some of the models. W.B.D. was supported by an NSERC Alexander 
Graham Bell Canada Graduate Scholarship. S.B. was supported by an NSERC Discovery Grant. M.W.K. was supported by STFC grant ST/F002505/2 
during the early phases of this work and is currently supported by the National Aeronautics and Space Administration through Einstein 
Postdoctoral Fellowship Award Number PF1-120084 issued by the Chandra X-ray Observatory Center, which is operated by the Smithsonian 
Astrophysical Observatory for and on behalf of the National Aeronautics Space Administration under contract NAS8-03060.

\bibliographystyle{aa}
\bibliography{bibliography}

\appendix

\section{Collision time scales}\label{app:collision_times}

We compute the collision times between the different species $s$ and neutrals according to the formula \citep[see][]{Mouschovias1996}
\begin{equation}
  \tau _{s\mathrm{n}} = k_{s,\mathrm{He}}\frac{m_{s}+m_{\mathrm{H}_{2}}}{\varrho_{\mathrm{n}}\langle \sigma w \rangle_{s\mathrm{H}_{2}}}.
\label{eq:collision_times}
\end{equation}
The quantity $k_{s,\mathrm{He}}$ is a correction factor entering the equation due to the fact that the gas also contains helium. 
The above expression hence calculates the collision time for a charged species $s$ with \textit{all} neutrals. Helium contributes only a small 
correction factor due to its low polarizability compared with $\mathrm{H}_{2}$ \citep[see][]{Spitzer1978}. 
\citet{Mouschovias1996} gives these correction factors as
\begin{equation}
k_{s,\mathrm{He}}=
  \begin{cases}
  1.23, & \text{if }s=i,\\
  1.21, & \text{if }s=e,\\
	1.09, & \text{if }s=g_{+}\text{, } g_{0} \text{, or } g_{-}.
\end{cases}
\label{eq:corr_factors_helium}
\end{equation}
The values for the rate constant $\langle \sigma w \rangle_{s\mathrm{H}_{2}}$ are taken from \citet{McDanielMason1973}, \citet{MottMassey1965}, and \citet{CiolekMouschovias1993}, respectively:
\begin{equation}
\langle \sigma w \rangle_{s\mathrm{H}_{2}}=
  \begin{cases}  
	1.69\times 10^{-9}~\mathrm{cm}^{3}~\mathrm{s}^{-1},  & \text{if }s=i,\\
  1.3\times 10^{-9}~\mathrm{cm}^{3}~\mathrm{s}^{-1}, & \text{if }s=e,\\
	\pi a^{2} \left(\left.8 k_{\mathrm{B}}T\right/\pi m_{\mathrm{H}_{2}}\right)^{1/2}, & \text{if }s=g_{+}\text{, } g_{0} \text{, or } g_{-}.
\end{cases}
\label{eq:rate_constants}
\end{equation}
The last expression assumes that the difference between the grain and neutral velocities is smaller than the sound speed.

\section{Rate Coefficients}\label{app:rate_coeffs}

For convenience, we reproduce here the rate coefficients used in this work. 
They can also be found in Appendix A of \citet{KunzMouschovias2009}. 

For radiative recombination of atomic ions and electrons, and for the dissociative 
recombination of electrons and $\mathrm{HCO}^{+}$ ions, we respectively adopt the 
values \citep{UmebayashiNakano1990}
\begin{align}
    \alpha_{\mathrm{rr}} &= 2.8 \times 10^{-12}~\left(300~\mathrm{K}/T \right)^{0.86}~\mathrm{cm}^{3}~\mathrm{s}^{-1},\\
    \alpha_{\mathrm{dr}} &= 2.0 \times 10^{-7}~\left(300~\mathrm{K}/T \right)^{0.75}~\mathrm{cm}^{3}~\mathrm{s}^{-1}.
\label{eq:alpha_rr_dr}
\end{align}
For charge-exchange reactions between atomic and molecular ions, we use the value from \citet{Watson1976}
\begin{equation}
    \beta = 2.5 \times 10^{-9}~\mathrm{cm}^{3}~\mathrm{s}^{-1}.
\label{eq:beta}
\end{equation}
The rate coefficients pertaining to ions (both molecular and atomic, indicated with subscript `i') and electrons (subscript `e') on the one hand and grains on the other are are taken from \citet{Spitzer1941,Spitzer1948}, with refinements made by \citet{DraineSutin1987} to account for the polarization of grains:
\begin{align}
    \alpha_{\mathrm{e}\mathrm{g}^{0}} &= \pi a^{2} \left(\frac{8k_\mathrm{B}T}{\pi m_{\mathrm{e}}} \right)^{1/2}
		                                     \left[1+ \left(\frac{\pi e^{2}}{2ak_\mathrm{B}T}\right)^{1/2}  \right] \mathcal{P}_{\mathrm{e}}, \\
    \alpha_{\mathrm{i}\mathrm{g}^{0}} &= \pi a^{2} \left(\frac{8k_\mathrm{B}T}{\pi m_{\mathrm{i}}} \right)^{1/2}
		                                     \left[1+ \left(\frac{\pi e^{2}}{2ak_\mathrm{B}T}\right)^{1/2}  \right] \mathcal{P}_{\mathrm{i}}, \\    
    \alpha_{\mathrm{e}\mathrm{g}^{+}} &= \pi a^{2} \left(\frac{8k_\mathrm{B}T}{\pi m_{\mathrm{e}}} \right)^{1/2} 
		                                     \left[1+ \left(\frac{e^{2}}{ak_\mathrm{B}T}\right)  \right]  \notag \\
																	  	&\times \left[1+ \left(\frac{2 e^{2}}{2e^{2}+ak_\mathrm{B}T}\right)^{1/2}  \right] \mathcal{P}_{\mathrm{e}}, \\
		\alpha_{\mathrm{i}\mathrm{g}^{-}} &= \pi a^{2} \left(\frac{8k_\mathrm{B}T}{\pi m_{\mathrm{i}}} \right)^{1/2}
				                                 \left[1+ \left(\frac{e^{2}}{ak_\mathrm{B}T}\right)  \right] \notag \\
		                                  &\times \left[1+ \left(\frac{2 e^{2}}{2e^{2}+ak_\mathrm{B}T}\right)^{1/2}  \right] \mathcal{P}_{\mathrm{i}}.
\label{eq:alpha_ei_g}
\end{align}
For the sticking probabilities of electrons or ions onto grains, we take the values from \citet{Umebayashi1983}, $\mathcal{P}_{\mathrm{e}}=0.6$ and $\mathcal{P}_{\mathrm{i}}=1.0$. In these equations, $a$ is the adopted grain radius, while other quantities have their usual meanings.
Lastly, the rate coefficients for charge transfer during collisions between grains are of the same form as the ones above, but with modified masses.
\begin{align}
    \alpha_{\mathrm{g}^{+}\mathrm{g}^{-}} &= 16\pi a^{2} \left(\frac{k_\mathrm{B}T}{\pi m_{\mathrm{g}}} \right)^{1/2}
				                                 \left[1+ \left(\frac{e^{2}}{2ak_\mathrm{B}T}\right)  \right] \notag \\
		                                     &\times \left[1+ \left(\frac{e^{2}}{e^{2}+ak_\mathrm{B}T}\right)^{1/2}  \right],\label{eq:alpha_gg}\\    
		\alpha_{\mathrm{g}^{\pm}\mathrm{g}^{0}} &= 16\pi a^{2} \left(\frac{k_\mathrm{B}T}{\pi m_{\mathrm{g}}} \right)^{1/2}
				                                 \left[1+ \left(\frac{\pi e^{2}}{4ak_\mathrm{B}T}\right)^{1/2}  \right] \mathcal{P}_{\mathrm{gg}}.
    \label{eq:alpha_gg0}
\end{align}
Here, $m_{\mathrm{g}}$ is the grain mass (assumed constant), and $\mathcal{P}_{\mathrm{gg}}\equiv 1/2$ is the probability of charge exchange between neutral and charged grains (both positive and negative). The probability of neutralization in Eq. (\ref{eq:alpha_gg}) is assumed to be unity.
\end{document}